\documentclass[aapm,mph,reprint,graphicx,nofootinbib,superscriptaddress,floatfix]{revtex4-1}
\usepackage[T1]{fontenc}
\usepackage{color,graphicx,amsmath}
\usepackage{amsmath,dcolumn,bm,siunitx}
\usepackage{graphicx,color}
\usepackage{physics}
\usepackage{siunitx}
\usepackage{placeins}
\usepackage[mathlines]{lineno}
\modulolinenumbers[5]
\usepackage[caption=false]{subfig}
\usepackage{adjustbox}
\usepackage{etoolbox}
\makeatletter
\patchcmd\linenumberpar{\@LN@parpgbrk}{\penalty\@LN@parpgpen\relax}{}{}
\makeatother
\graphicspath{{figures/pdf/}}

\def\SupSecA{Sup.~Sec.~A}       
\def\SupSecB{Sup.~Sec.~B}       
\def\SupTableII{Sup.~Table~II}  
\def\SupTableIII{Sup.~Table~III}
\def\SupFig#1{Sup.~Fig.~#1}

\begin{document}
\title{Design and Characterization of Tissue-Mimicking Gel Phantoms for Diffusion Kurtosis Imaging}
\author{Ziyafer Gizem Portakal} 
\affiliation{Department of Physics, Science and Arts Faculty, Cukurova University, 01330, Adana, Turkey}
\affiliation{Department of Medical Physics, Velindre Cancer Centre, CF14 2TL, Cardiff, United Kingdom}
\author{Sophie Shermer}
\affiliation{Department of Physics, College of Science, Swansea University, SA2 8PP, Swansea, United Kingdom}
\email{lw1660@gmail.com}
\author{Christopher Jenkins}
\affiliation{Department of Physics, College of Science, Swansea University, SA2 8PP, Swansea, United Kingdom}
\affiliation{Institute of Life Science, College of Medicine, Swansea University, Swansea SA2 8PP, United Kingdom}
\author{Emiliano Spezi} 
\affiliation{Department of Medical Physics, Velindre Cancer Centre, CF14 2TL, Cardiff, United Kingdom}
\affiliation{School of Engineering, Cardiff University, CF24 3AA, Cardiff, United Kingdom}
\author{Teresa Perrett} 
\affiliation{Department of Medical Physics, Velindre Cancer Centre, CF14 2TL, Cardiff, United Kingdom}
\author{Nina Tuncel} 
\affiliation{Department of Physics, Science Faculty, Akdeniz University, 07058, Antalya, Turkey}
\author{Jonathan Phillips}
\affiliation{Institute of Life Science, Medical School, Swansea University, Swansea SA2 8PP, United Kingdom}
\date{\today}

\begin{abstract}
\textbf{Purpose:}
The aim of this work was to create tissue-mimicking gel phantoms appropriate for diffusion kurtosis imaging (DKI) for quality assurance, protocol optimization and sequence development. 

\textbf{Methods:} A range of agar, agarose and polyvinyl alcohol phantoms with concentrations ranging from $1.0\%$ to $3.5\%$, $0.5\%$ to $3.0\%$ and 10\% to 20\%, respectively, and up to \SI{3}{g} of glass microspheres per \SI{100}{ml} were created.  Diffusion coefficients, excess kurtosis values and relaxation rates were experimentally determined.

\textbf{Results:} The kurtosis values for the plain gels ranged from $0.05$ with 95\% confidence interval (CI) of $(0.029,0.071)$ to $0.216(0.185,0.246)$, well below the kurtosis values reported in the literature for various tissues. The addition of glass microspheres increased the kurtosis of the gels with values up to $0.523(0.465,0.581)$ observed for gels with the highest concentration of microspheres.  Repeat scans of some of the gels after more than six months of storage at room temperature indicate changes in the diffusion parameters of less than 10\%.  The addition of the glass microspheres reduces the apparent diffusion coefficients (ADCs) and increases the longitudinal and transverse relaxation rates but the values remain comparable to those for plain gels and tissue, with ADCs observed ranging from $818 (585,1053) \times 10^{-6}\si{mm^2/s}$ to $2257(2118,2296) \times 10^{-6}\si{mm^2/s}$, and $R_1$ values ranging from $0.34(0.32,0.35)$ \si{1/s} to $0.51 (0.50,0.52)$ \si{1/s}, and $R_2$ values ranging from $9.69(9.34, 10.04)$ \si{1/s} to $33.07(27.10, 39.04)$ \si{1/s}.

\textbf{Conclusions:} Glass microspheres can be used to effectively modify diffusion properties of gel phantoms and achieve a range of kurtosis values comparable to those reported for a variety of tissues.
\end{abstract}

\maketitle

\section{Introduction}

Diffusion-weighted magnetic resonance imaging (DW-MRI) is an important non-invasive technique to obtain information about tissue microstructure via quantities such as the apparent diffusion coefficient (ADC) or fractional anisotropy (FA) in the more general case of diffusion tensor imaging (DTI)~\cite{Basser1994, Bihan2013}. By acquiring a series of images with varying degrees of diffusion weighting, parametric maps can be computed allowing qualitative and quantitative assessment of the diffusion behavior.  The simplest model for water diffusion in tissue is a Gaussian random process~\cite{Tanner1965}, leading to linear decay of the natural logarithm of the DW-MRI signal intensity with increasing degrees of diffusion weighting.  However, due to the complex structure of most biological tissues, the diffusion displacement probability distribution can deviate substantially from a Gaussian form~\cite{Karger1985}. Diffusion kurtosis imaging (DKI) aims to capture the degree to which such diffusion processes are non-Gaussian by replacing the mono-exponential fitting of the signal by a quadratic one with a (dimensionless) coefficient $K$ quantifying the (excess) kurtosis, or simply the kurtosis. DKI was first suggested in~\cite{Jensen2005} and has since been applied \textit{in vivo} in many situations \textit{e.g.} grading of cerebral gliomas~\cite{Raab2010a}, head and neck squamous cell carcinoma~\cite{Raab2010b}, prostate cancer~\cite{Rosenkrantz2012,Rosenkrantz2013,Quentin2014,Tamura2014,Suo2014}, breast cancer~\cite{Nogueira2014} and even in the lungs~\cite{Trampel2006} using hyperpolarized $^{3}$He.

As with all quantitative imaging modalities, the development and validation of imaging protocols and pulse sequences in the case of MRI are of utmost importance. This necessitates the development of experimentally well-characterized tissue-mimicking phantoms that can be used repeatedly for calibration, development, testing, quality assurance and indeed understanding the physics governing the imaging observations. Phantoms can be tissue-mimicking in various aspects.  The relevant properties depend on the application.  In general, proton density, homogeneity on a desired length scale and relaxation times similar to those of the tissues being modeled are desirable for MRI phantoms.  Room-temperature DKI phantoms must furthermore exhibit diffusion properties at room temperature comparable to those of healthy or diseased tissue at body temperature.

The main types of phantoms used in MRI studies are aqueous solutions and gels.  Although easy to prepare, aqueous solutions are not tissue-mimicking in many regards, having $T_{1}$ (spin-lattice) and $T_{2}$ (spin-spin) relaxation times that are approximately equal, unlike human tissue, for which the $T_{2}$ values are typically much shorter than $T_{1}$.  Gel phantoms, on the other hand, can be prepared to mimic the $T_{1}$ and $T_{2}$ values of human tissues~\cite{Hattori2013}. They are also relatively easy to make and use, cost-effective, reusable, less prone to leakage and have a long lifetime when a preservative material is added to inhibit bacterial growth.  Aside from synthetic gelling agents such as polyvinyl alcohol (PVA), the most common source of biological gelling agents such as agar, agarose and carageenan are seaweeds and algae.

Water-based tissue-mimicking gels made from agar, agarose or PVA have been investigated extensively in the literature, including various types of phantoms for the assessment of diffusion~\cite{Laubach1998,Lavdas2013} and anisotropic diffusion~\cite{Fieremans2008, Farrher2012}.  For the assessment of kurtosis three test objects have been reported: homogenized asparagus~\cite{Jensen2005}, dairy cream~\cite{Fieremans2012} and colloidal dispersions~\cite{Phillips2015}.  Homogenized asparagus and cream were found to have a mean (directionally averaged) kurtosis value of $K_\textrm{asp}=0.28\pm0.05$ and $K_{\textrm{cream}}=1.18\pm 0.04$, respectively.  Whilst displaying kurtosis values in the range reported in vivo (see Table~\ref{table:kurtosis}) both the asparagus and the cream phantoms, though very useful, are perishable and the composition of the phantoms may vary between samples, and therefore not ideal for use as long term test objects. The colloidal dispersions reported in \cite{Phillips2015} displayed kurtosis values in the range $0 \leq K_{\textrm{coll}} \leq 0.62$ and while suitable for long term use, they do not possess the same relaxation rates as tissue, which would be advantageous for multi-modality imaging. 

It has been claimed that kurtosis is related to barrier concentration \textit{e.g.}~\cite{Novikov2011,Chu-Lee2013,Phillips2015} and also $R_{2}^*$~\cite{Palombo2015} although the fundamental origin of non-Gaussian diffusion in biological systems is not fully understood.  It is not our aim in this paper to investigate the microscopic origin of the diffusion-weighted MR signal but to create tissue-mimicking gel phantoms appropriate for DKI with relaxation rates similar to tissue by characterizing the ADCs, kurtosis values and relaxation times of agar, agarose and PVA phantoms. We focus on creating isotropic kurtosis phantoms as a starting point for creating more complex phantoms that can model complex anisotropic diffusion in biological tissues such as brain~\cite{Armitage1998,Shimony1999, Beaulieu2002, Oouchi2007} and prostate~\cite{Reinsberg2005,Xu2009,Ogura2011}.  Both pure gels and gels with various concentrations of additives such as glass microspheres are characterized in terms of their relaxation rates and diffusion properties, including kurtosis.  The gels considered are liquids immersed in a macromolecular framework. As such the diffusion of the water molecules should be hindered by the presence of the macromolecular skeleton. The addition of glass microspheres further increases the barrier concentration and hence the non-Gaussian behavior of the diffusion process should increase with the addition of such glass microspheres. 

\section{Methods}

\subsection{Phantom Preparation}

\begin{figure}[h]
  \includegraphics[width=0.40\textwidth]{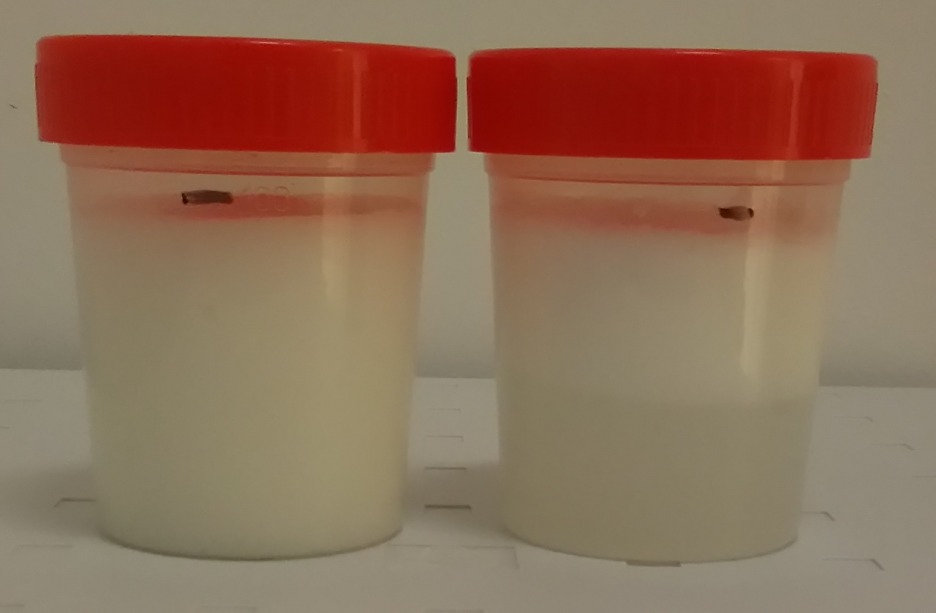}
  \caption{Good gel phantoms have a uniform appearance (left) while failed attempts for which the microspheres have separated before solidification show distinct layers (right).}
\label{fig:Phantoms}  
\end{figure}

Multiple gel phantoms were prepared using different gelling agents including agar (\#A7002, Sigma-Aldrich, Dorset, UK), agarose (\#A0169, Sigma-Aldrich, Dorset, UK) and PVA (99+\% hydrolysis degree, \#363146, Sigma-Aldrich, Dorset, UK) at the  Cancer Research Wales Laboratories in Velindre Cancer Centre. Homogeneous gel phantoms were created at different concentrations using desired powders dissolved in \SI{18.2}{\mega\ohm \cdot\centi\meter} distilled water and heated up to \SIrange{80}{90}{^\circ C} while mixing for 30-45 minutes. Diazolidinyl urea (DU) (\#D5146, Sigma-Aldrich, Dorset, UK) was added into the mixtures at 6 mg per ml to prevent bacterial growth. Solidification and polymerization of plain agar and agarose gels occurs overnight at room temperature. The process can be accelerated by refrigeration or immersion in ice water. For PVA cryogels, gelation is induced by freeze-thaw (FT) cycles, which involve placing the PVA phantoms in a freezer at \SI{-20}{^\circ C} for 10 hours and then leaving them at room temperature (\SI{20}{^\circ C}) for 14 hours. For this work four freeze-thaw cycles were used.  For the gels with glass microspheres varying amounts of glass microspheres (\#K20, 3M microspheres, Easy Composites Ltd., Staffordshire, UK) with diameters ranging from \SI{30}{\micro\meter} to \SI{90}{\micro\meter} were added. Fifteen plain gel phantoms (six agar, six agarose, three PVA) and 22 gel phantoms containing varying concentrations of glass microspheres (fourteen agar, eight agarose) were created.  Each phantom has a volume of \SI{100}{ml} and is approximately \SI{5}{cm} in diameter and \SI{5}{cm} in height.  The gels are stored in containers made of high density polyethylene (HDPE) with tightly sealing lids.  Attempts to create PVA cryogels with the chosen microsphere material were abandoned as the glass microspheres appeared to react with the PVA cryogels, resulting in a sticky gum-like material. For simplicity we shall refer to a gel with $x$ \si{g} agar/agarose/PVA per \SI{100}{ml} as an $x$ \% agar/agarose/PVA gel.  When adding microspheres that have a tendency to float to the top, care must be taken to ensure the microspheres are properly blended with the gels and that gelification occurs fast enough to prevent the separation of the microspheres.  Ideally, the additive would be density-matched to the surrounding material to avoid a sedimentation effect but we found that good results could still be achieved with the microspheres used, provided the gels were carefully prepared.   Furthermore, separation of microspheres can usually be detected by visual inspection of the gels after solidification (see Fig.~\ref{fig:Phantoms}).

\subsection{DW-MRI Measurements}

\begin{figure*} 
  \subfloat[Agar 1\%, $b=0$]{\includegraphics[width=0.23\textwidth]{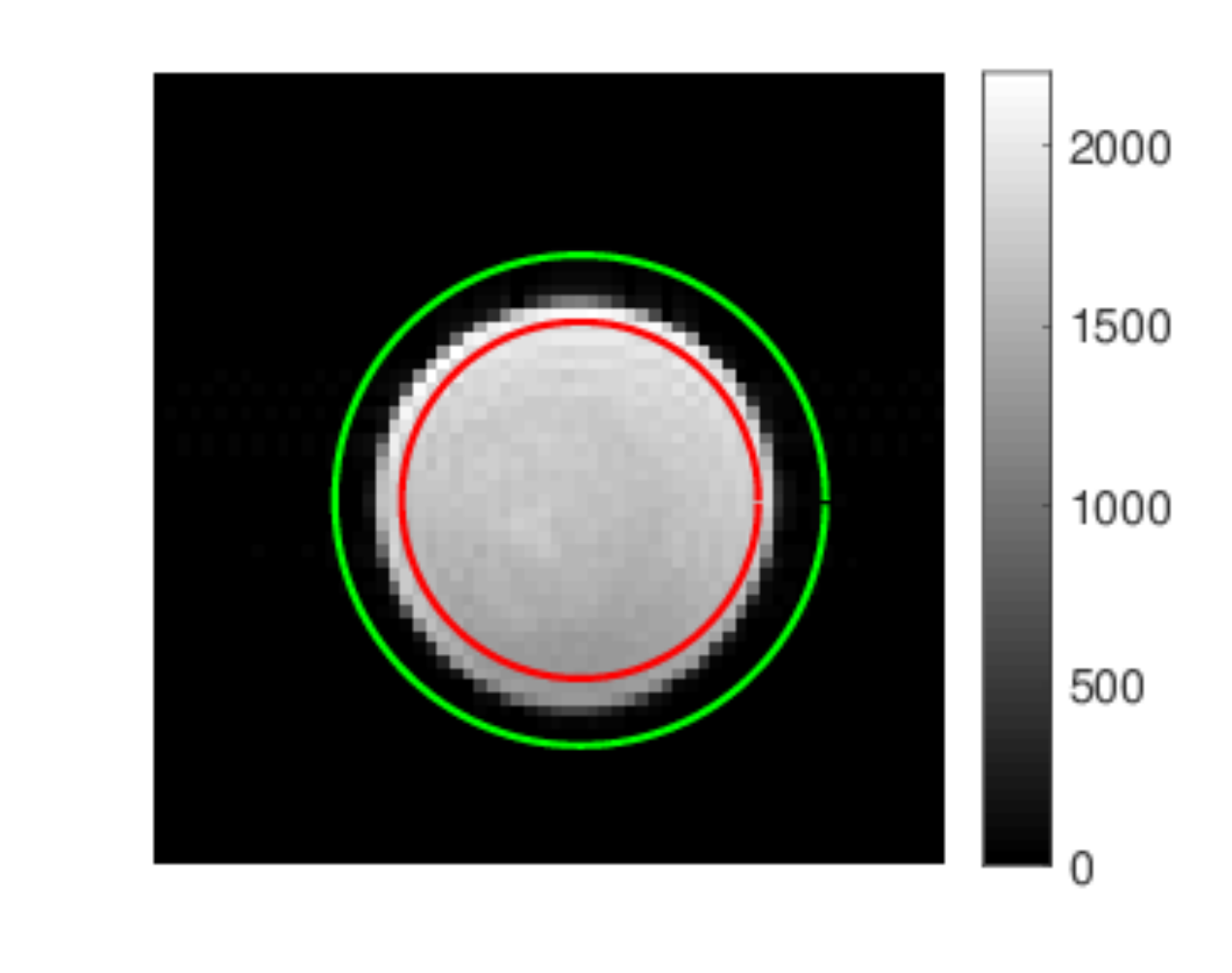}\hspace{-0.05in}}
  \subfloat[Agar 1\%, $b=4000$]{\includegraphics[width=0.23\textwidth]{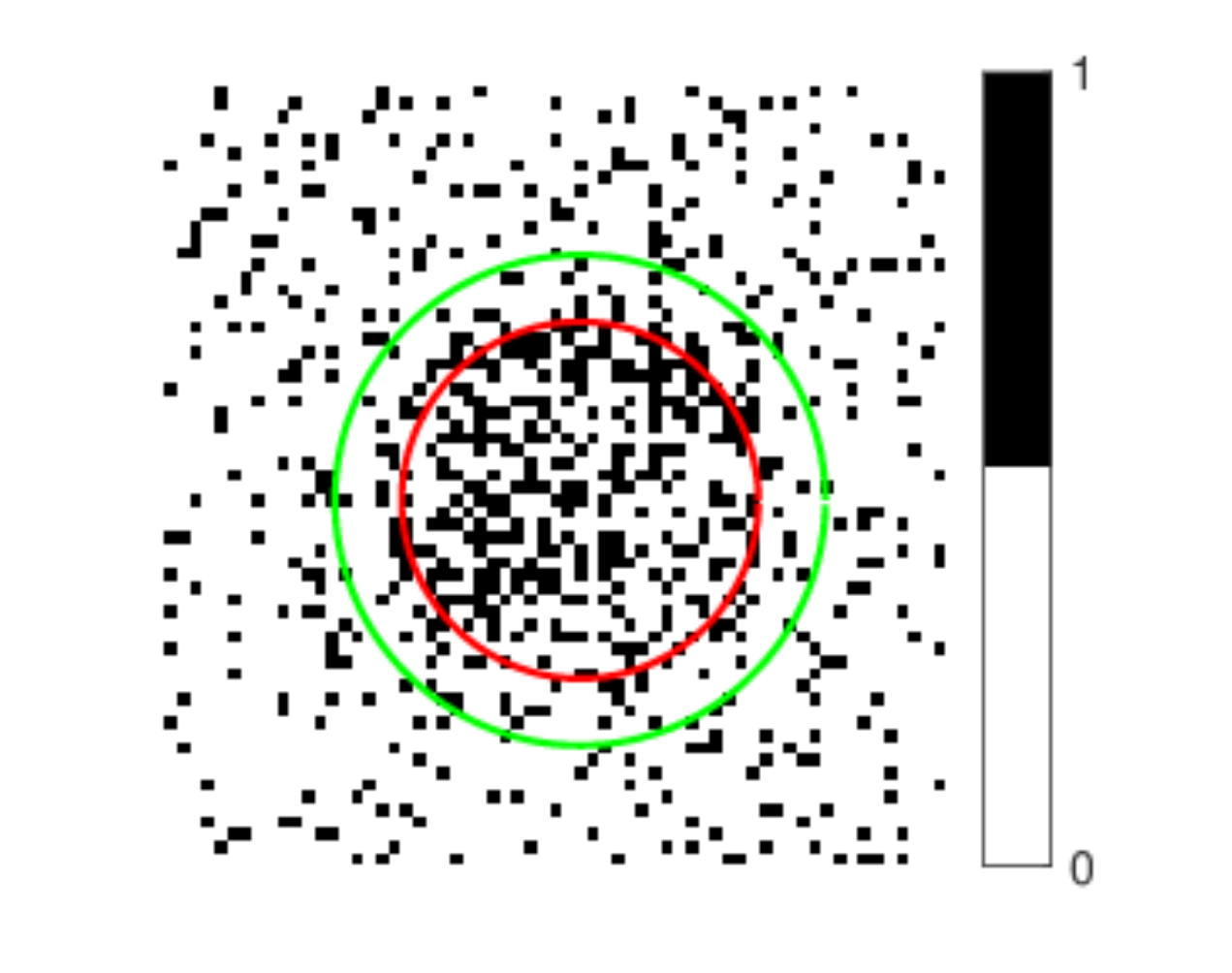}\hspace{-0.05in}}
  \subfloat[Agar 2\%, \SI{3}{g}, $b=0$]{\includegraphics[width=0.23\textwidth]{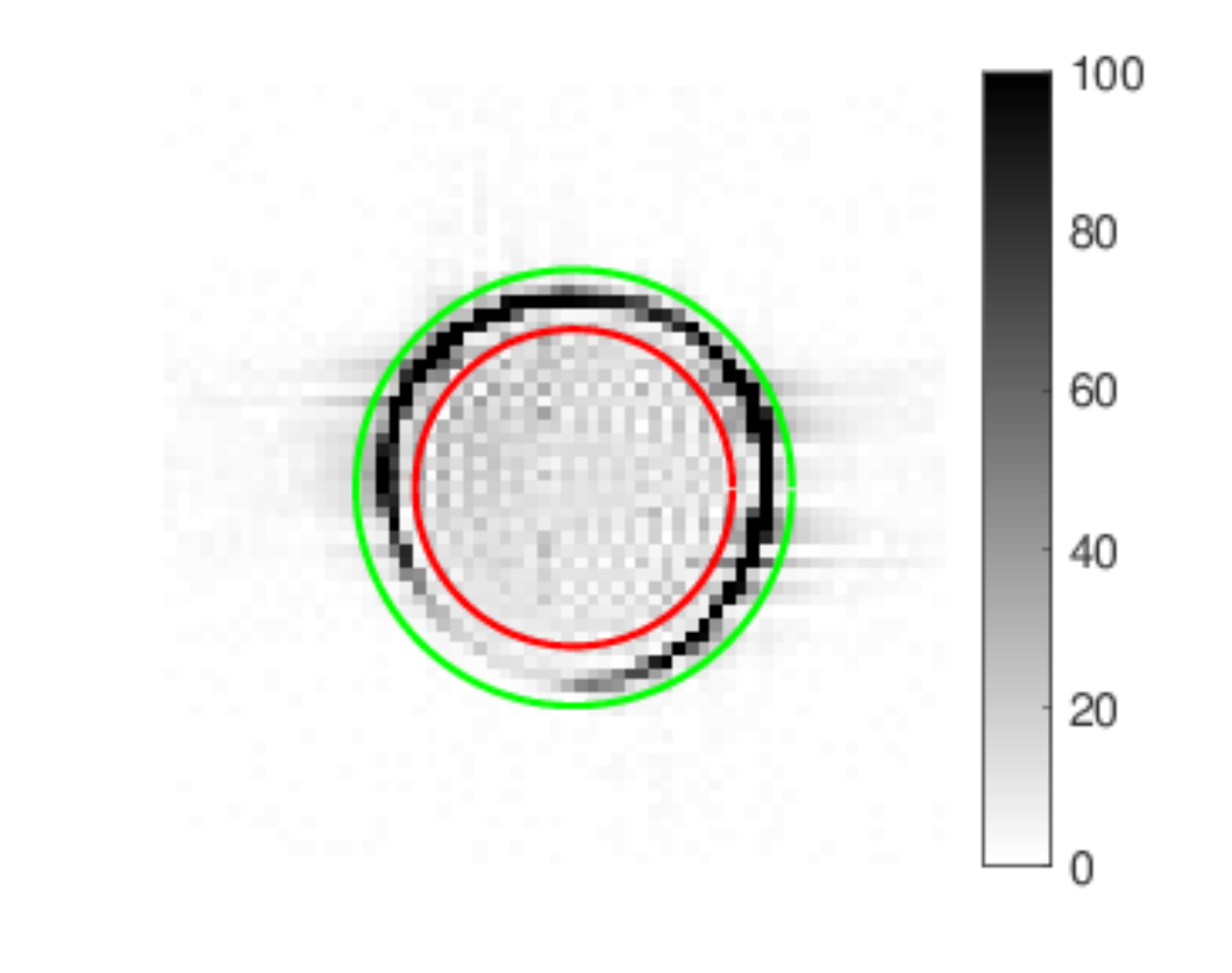}\hspace{-0.1in}}
  \subfloat[Agar 2\%, \SI{3}{g}, $b=4000$]{\includegraphics[width=0.23\textwidth]{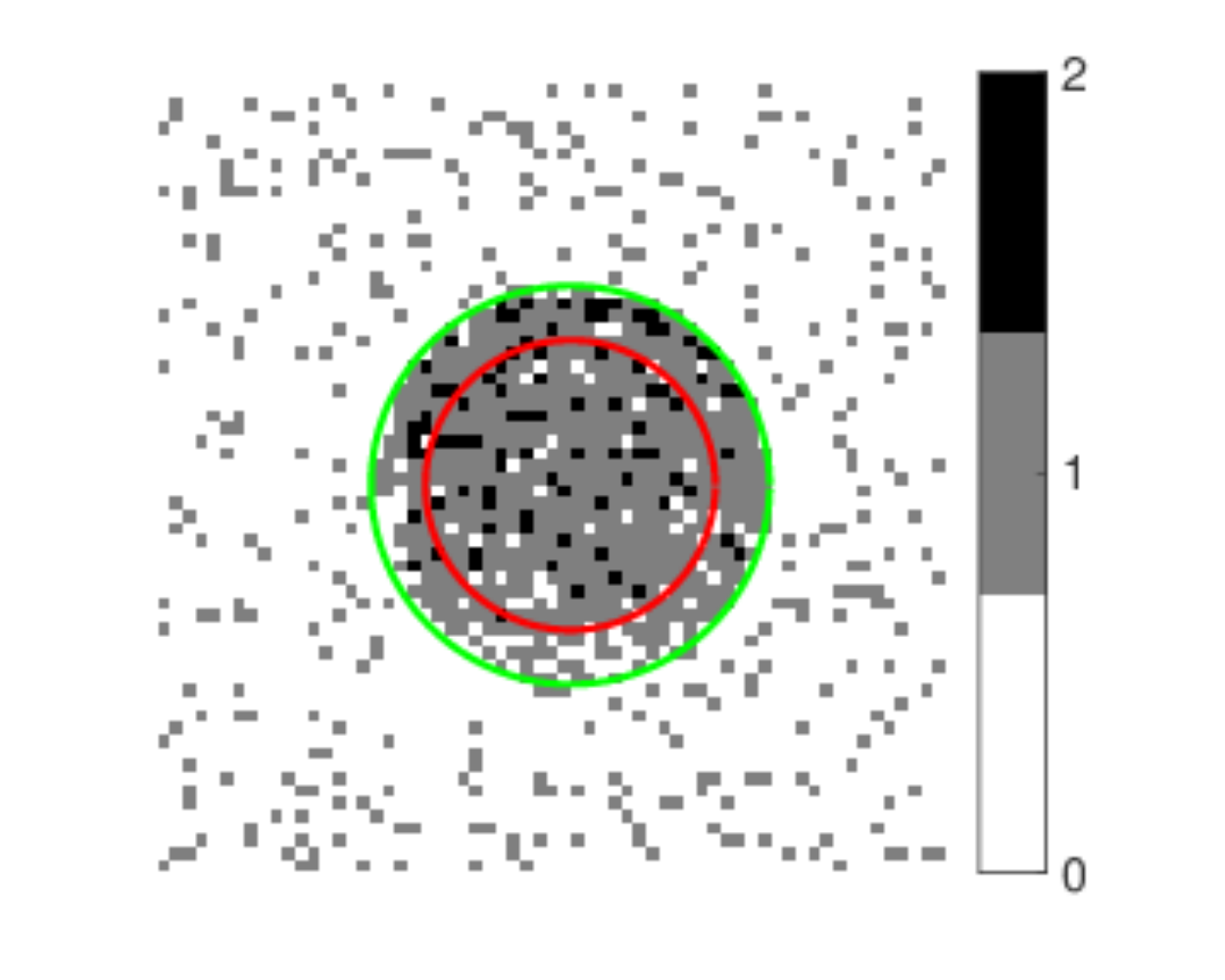}}

  \caption{Images and ROI selection for a 1\% plain agar phantom and a 2\% agar phantom with \SI{3}{g} of microspheres for $b=\SI{0}{s/mm^2}$ and $b=\SI{4000}{s/mm^2}$.  The inside of the red circle (ROI1) covers at least 75\% of the phantom area while the background, defined by the outside of the green circle, ROI2, is as large as possible while avoiding overlap with the phantom, container or artifacts. All images were acquired for a field-of-view (FOV) of $\SI{100}{mm}\times \SI{100}{mm}$ and matrix size $64\times 64$.}
\label{fig:images}
\end{figure*}

DWI MRI data were acquired on a Siemens 3T Magnetom Skyra (Erlangen, Germany) scanner at Swansea University using a combination of a \SI{7}{cm} diameter loop and a four-channel spine coil element (SP2) to boost the diffusion-weighted signal. The signal to noise ratio (SNR) of DW-MRI scans was tested using different single coils as well as dual coil combinations.  The loop and spine coil combination was chosen as it provided the best results to calculate the diffusive properties~\cite{Mitsouras2004}.  All scans were performed at room temperature in an air-conditioned and temperature-controlled environment at $20^\circ \pm 0.6^\circ$C.

Diffusion-weighted images were acquired using an in-house version of a spin-echo sequence with a pair of strong diffusion-weighting gradients straddling the 180$^\circ$ refocusing pulse known as the Stejskal-Tanner or pulsed gradient spin echo (PGSE) sequence~\cite{Tanner1968}.  The sequence was written in the Siemens \texttt{IDEA} \texttt{C++} programming environment.  For the purposes of this investigation, data quality was prioritized over rapid acquisition.  Therefore, the scans were performed using normal Cartesian $k$-space readout rather than single-shot echo planar imaging (SS-EPI) to avoid effects such as image blurring, localized signal loss, image distortions caused by eddy currents~\cite{Farzaneh1990,Bammer2009} and other artifacts observed with the standard EPI diffusion sequence.  Our custom PGSE sequence was tested with a water phantom as a negative control and compared to the vendor-supplied diffusion sequence.  The results are provided in \SupSecA.  The parameters for the PGSE phantom scans were: FOV \SI{100}{mm} $\times$ \SI{100}{mm}, matrix size $64\times 64$, $T_{E}= \SI{120}{ms}$, $T_{R}=\SI{3000}{ms}$, readout bandwidth of \SI{130}{Hz/pixel} and slice thickness \SI{20}{mm}.

In-house software written in \texttt{MATLAB} was used to analyze the images, including automatic region-of-interest (ROI) selection as well as mean and pixel-by-pixel signal analysis.  Thresholding was used to locate the sample and select a circular ROI about 80\% of the diameter of the sample (ROI1).  80\% of the maximum diameter of the sample was chosen instead of 90\% to avoid the interface region with the container.  It is important to stay clear of the noise floor to avoid artificially inflated kurtosis estimates due to pseudo-kurtosis.  In the following, phantom-air contrast, where the ``air'' ROI (ROI2) was defined as all voxels outside a circle approximately $1.2$ times the diameter of the sample, as shown in Fig.~\ref{fig:images}, was used to select the $b$-values to be included in each ADC/kurtosis fit.  A detailed analysis of SNR, definition of phantom-air contrast and comparison of different methods of assessing noise are provided in \SupSecB.
 
The ADC is calculated assuming linear decay of the natural logarithm of the DW-MRI signal intensity ($I$) with increasing $b$-value
\begin{equation}
\label{linear_model}
\log \left( \frac{I}{I_{0}}  \right) = - bD,
\end{equation}
where $I_{0}$ is the signal intensity at $b=0$, $D$ is the ADC, typically measured in \si{mm^2/\second} and  $b$ is the degree of diffusion weighting, typically measured in \si{s/mm^2}, given by
\begin{equation*}
b = (\gamma g \delta)^{2} \left( \Delta -\delta/3  \right),
\end{equation*}
neglecting terms arising from the rise and decay times of the gradients, which are negligible in our case.  $\gamma$ is the gyromagnetic ratio, $g$ is the amplitude of the diffusion-encoding gradient pulse in \si{T/m}, $\delta$ is the duration of a single diffusion gradient in \si{ms} and $\Delta$ the delay between the gradients in \si{ms}.  For trapezoidal gradient pulses with rise (fall) time $t_{\rm rise}$ ($t_{\rm fall}$) and flat-top time $t_{\rm flat}$ the \emph{duration} $\delta$ is defined here as $t_{\rm rise} + t_{\rm flat}$ instead of the total pulse length $t_{\rm rise}+t_{\rm flat}+t_{\rm fall}$, so that the gradient moment for a symmetric trapezoidal gradient with $t_{\rm fall}=t_{\rm rise}$ is $\delta$ times the gradient amplitude $g$.  The gradient parameters used were similar to those used in the vendor supplied product sequence, specifically $\delta=\SI{33000}{\micro\second}$, $\Delta=\SI{71000}{\micro\second}$ and $t_{\rm rise}=\SI{500}{\micro\second}$, and the gradient amplitudes calculated according to the $b$-value required.

The kurtosis, a measure of the deviation of the diffusion propagator from a Gaussian form, was estimated using the quadratic exponential model 
\begin{equation}
  \label{quad_model}
\log \left( \frac{I}{I_{0}}  \right) = -bD +\frac{1}{6}b^{2}D^{2}K +\mathcal{O}(b^{3}),
\end{equation}
where $D$ is the diffusion coefficient and $K$ the dimensionless (excess) kurtosis.  To distinguish the diffusion coefficients for the linear and quadratic model we use $D^{(1)}$ to denote the diffusion coefficient using the standard linear model~\eqref{linear_model} or ADC, and $D^{(2)}$ to denote the diffusion coefficient using the quadratic model~\eqref{quad_model}.  In addition to mean signal analysis, voxel-based linear and quadratic fits were performed to map the spatial variation of the diffusion parameters for each phantom.

\subsection{MRI Relaxation Rate Measurements}

Relaxation rate measurements were performed on the same \SI{3}{T} scanner using a four-channel spine coil element (SP2).  Spin and gradient echo sequences with different echo and repetition times, $T_{E}$  and $T_{R}$, respectively, were used to evaluate the relaxation properties of the phantoms.  $R_{1}$ ($1/T_1$)  was determined using a saturation recovery protocol. This involved repeated scans with a vendor-supplied spin echo sequence, consisting of a $90^\circ$ excitation pulse and a $180^\circ$ refocusing pulse, with a fixed $T_E$ of \SI{12}{ms} and different $T_{R}$ of 125, 250, 500, 1000, 2000, 3000, 4000, 6000 and \SI{7000}{ms}.  $R_{2}$ ($1/T_2$) was determined using a vendor-supplied multi-spin echo sequence with $T_{E}=n \cdot \SI{15}{ms}$ for $n$ ranging from $1$ to $32$ and a long $T_{R}$ of \SI{6000}{ms} to ensure that the longitudinal magnetization recovers sufficiently to avoid stimulated-echo effects.  $R_2^*$ was determined by acquiring a series of images using a vendor-supplied gradient echo sequence with fixed $T_R=\SI{3000}{ms}$ and different echo times $T_E$ ranging between \SI{2}{ms} and \SI{60}{ms} for a \SI{10}{mm} coronal slice through the center of the phantom.  For the $R_1$ and $R_2$ measurements the FOV was \SI{100}{mm} $\times$ \SI{100}{mm}, the matrix size $128\times 128$ and the readout bandwidth \SI{130}{Hz/Px}.  For the $R_2^*$ ($1/T_2^*$) measurements multiple gels were scanned simultaneously using a FOV of \SI{128}{mm} $\times$ \SI{128}{mm}, matrix size $128\times 128$ pixels and readout bandwidth of \SI{505}{Hz/Px}.

In-house software written in \texttt{MATLAB} was used to analyze the images.  ROI selection was performed as described above.  For the bulk analysis the mean and standard deviation of the signal over the ROI were determined for each image.  $R_{1}$ was determined by fitting the mean signal vs $T_{R}$  according to
\begin{equation}
   I(T_{R}) = I_{0} \left[ 1-\exp \left( -R_1 T_{R}  \right)   \right].
\end{equation}
$R_2^*$ was determined by fitting the mean signal over the ROI vs $T_E$ for a sequence of gradient echo-images acquired for different $T_E$ according to
\begin{equation}
  \log I(T_{E}) = - R_2^*  T_{E} +a_{0}^*.
\end{equation}
$R_2$ was determined by fitting the mean signal over the ROI vs $T_E$ for a series of spin echo images acquired using a multi-spin-echo sequence via
\begin{equation}
  \log I(T_{E}) = - R_2  T_{E} +a_{0}.
\end{equation}
The number of echoes used for the linear fit was adjusted to avoid noise floor issues for long $T_{E}$ for samples with large $R_2$.  For most samples 16 echoes were fitted.   The parameters $a_0=\log(I_0)$ and $a_0^*=\log(I_0^*)$, where $I_0$ and $I_0^*$ are constants related to the equilibrium magnetization, coil sensitivities and spin density of the sample, which are not used in the following.

\subsection{Repeat measurements and QA}

To assess the temporal stability of the phantoms, the diffusion and relaxation rate measurements were repeated for some of the phantoms after they had been stored at room temperature for more than six months.  For the diffusion scans the same scan sequence and parameters were used.  The repeat scans for the relaxation measurements were performed using the same sequences and parameters as detailed above, except a larger \SI{200}{mm} FOV to enable scanning multiple phantoms concurrently.

To further assess the spatial variation of $R_2^*$ in the $y$-direction, e.g., due to concentration gradients, a multi-slice $R_2^*$ protocol was added.  Using the same vendor-supplied gradient echo sequence used for previous $R_2^*$ scans, twenty \SI{2}{mm} coronal slices were acquired for each phantom for a fixed $T_R=\SI{500}{ms}$ and multiple $T_E$ ranging from \SI{3}{ms} to \SI{21}{ms} with a FOV of $\SI{100}{mm}\times \SI{100}{mm}$, matrix size $64\times 64$, and bandwidth of \SI{510}{Hz/px}.
  
To facilitate slice selection and avoid regions with high $B_0$ inhomogeneity, the spatial homogeneity of the $B_0$ field was investigated by obtaining double-echo interference field maps for all phantoms using a vendor-supplied service sequence with $T_R=\SI{300}{ms}$, $T_E=\SI{20}{ms}$, FOV $\SI{128}{mm} \times \SI{128}{mm}$, matrix size $128 \times 128$ and readout bandwidth of \SI{130}{Hz/px}. The sequence used is similar to a CPMG sequence but deliberately calibrated to generate both stimulated and regular spin echos with approximately equal strength.  In the presence $B_0$ inhomogeneity an interference pattern is observed.  Closely spaced dark fringes in a particular region indicate high $B_0$ inhomogeneity, while absence of fringes indicate good $B_0$ homogeneity.  For the chosen parameters the difference in the Lamor frequency between two pixels corresponding to adjacent dark fringes is \SI{50}{Hz}.

\section{Results}

In the following, a parameter with (most likely) value $x$ and 95\% confidence interval $(a,b)$ shall be denoted by $x(a,b)$.  Examples of \SI{100}{ml} phantoms resulting from the preparation process described are shown in Fig.~\ref{fig:Phantoms}.

\subsection{Diffusive Properties}

\begin{table}
  \caption{Kurtosis values reported for various tissues} \label{table:kurtosis}
  \begin{tabular}{cc}
    Lung, diseased       & 0.21 \cite{Trampel2006} \\
    Lung, healthy         & 0.34 \cite{Trampel2006} \\
    Grey matter           & 0.41 \cite{Minati2007} \\
    White matter          & 0.70 \cite{Minati2007} \\
    Prostate, healthy    & 0.57 \cite{Lawrence2012} \\
    Prostate, diseased & 1.05 \cite{Lawrence2012} 
  \end{tabular}
\end{table}

\begin{table*}
\scalebox{1.0}{\begin{minipage}{\textwidth}
\begin{tabular}{|l||c|c|c|} \hline
& $D^{(1)} (\SI{1e-6}{mm^2/s})$ &  $D^{(2)} (\SI{1e-6}{mm^2/s})$ & $K$ \\\hline\hline
 1.0\% Agar, scan 1                             & 2184 (2140,2227) & 2288 (2198,2377) & 0.050 (0.029,0.071)\\\hline
 1.0\% Agar, scan 2                             & 2061 (2037,2084) & 2123 (2070 2174) & 0.045 (0.030,0.059)\\\hline
 1.0\% Agarose, scan 1                       & 2259 (2168,2350) & 2344 (2283,2404) & 0.059 (0.046,0.072)\\\hline
 1.0\% Agarose, scan 2                       & 2114 (2057,2172) & 2143 (2100,2186) & 0.030 (0.018,0.043)\\\hline
 1.0\% Agarose, 1.0g spheres, scan 1 & 1962 (1865,2060) & 2227 (2125,2328) & 0.196 (0.183,0.209)\\\hline
 1.0\% Agarose, 1.0g spheres, scan 2 & 1997 (1906,2088) & 2281 (2167,2395) & 0.176 (0.161,0.192)\\\hline
\end{tabular}
\end{minipage}}
\caption{Diffusion and kurtosis parameters (with 95\% confidence intervals in parentheses) for scans more than 6 months apart suggest good long-term stability with only small decreases observed for most parameters.  The values for the gel with microspheres appear more stable than those for the plain gels.}
\label{table:3}
\end{table*}

The diffusion coefficient $D^{(1)}$ obtained from linear fits of the logarithm of the signal intensity (Eq.~\eqref{linear_model}) in the diffusion-weighted image data as well as the diffusion coefficient $D^{(2)}$ and kurtosis values obtained from the quadratic fit of the logarithm of the signal (Eq.~\eqref{quad_model}) of the data are shown in \SupTableIII.  For the linear fit only the first four $b$-values (\SIlist{0; 500; 1000;1500}{s/mm^2}) were used, while all nine $b$-values (\SIrange{0}{4000}{s/mm^2}) were used for the quadratic fit.  

\begin{figure}[h]
  \includegraphics[width=0.45\textwidth]{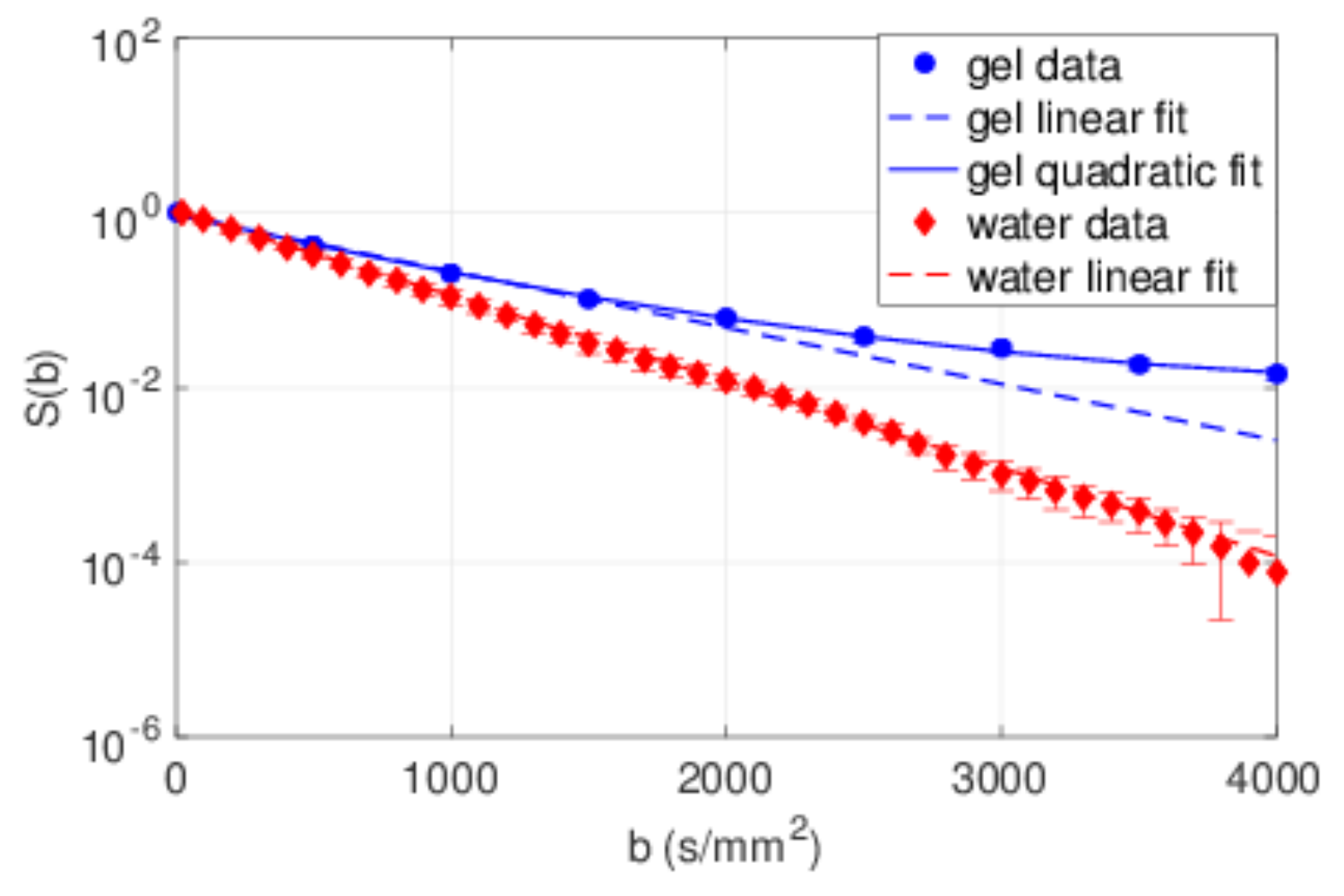}
  \caption{Mean signal (marker) and standard deviation of the signal (error bars) over phantom ROI (ROI1) as a function of the $b$-value as well as the linear fit (dashed line) and quadratic fit (solid line) for a 2\% agar phantom with \SI{0.1}{g} glass microspheres (blue) and a water phantom (red).}
\label{fig:diffusion-fit} 
\end{figure}

Fig.~\ref{fig:diffusion-fit} shows the signal, scaled so that $S(0)=1$, as a function of the $b$-value, for a 2\% agar gel with \SI{0.1}{g} microspheres and a water phantom control.   For the water phantom the logarithm of the signal decays linearly, as expected, with $D^{(1)} = 2254(2228,2280) \times 10^{-6}$~\si{mm^2/s} while the signal decay for the gel phantom, even for this low concentration of microspheres, is nonlinear and best described by a quadratic exponential fit with diffusion cofficient $D^{(2)}=2421(2217,2625) \times 10^{-6}$~\si{mm^2/s} and kurtosis of $0.177(0.154,0.200)$.

\begin{figure*}
\includegraphics[width=0.85\textwidth]{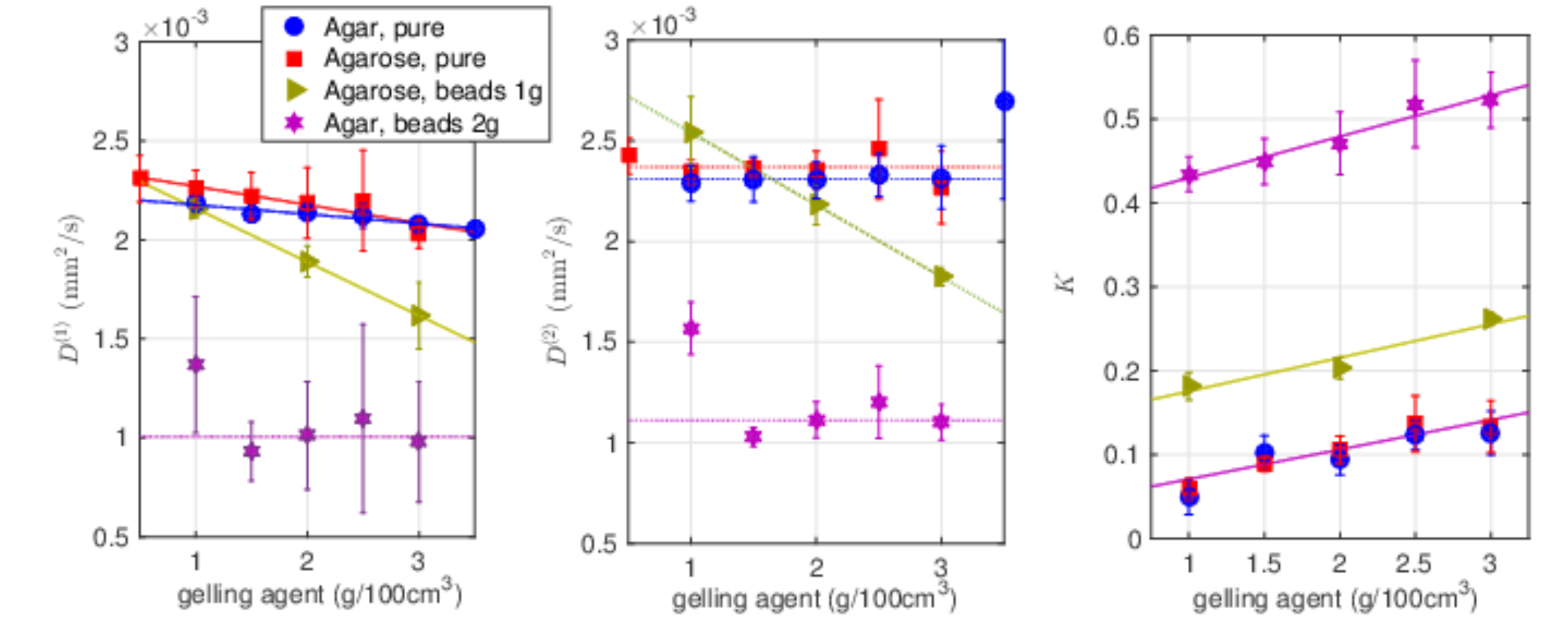}
\caption{Diffusion coefficients $D^{(1)}$ and $D^{(2)}$ and kurtosis $K$ as a function of the gelling agent concentration, with markers and error bars indicating mean values and 95\% confidence intervals obtained from the respective model fits for various pure agar and agarose gels (blue dots and red squares, respectively), as well as agar and agarose gels with microspheres (magenta stars and yellow triangles, respectively).  The dotted lines indicate trends determined by linear regression fits of the diffusion coefficients and kurtosis as a function of gelling agent concentration.} \label{fig:diffusion-vs-gel-conc}
\includegraphics[width=0.85\textwidth]{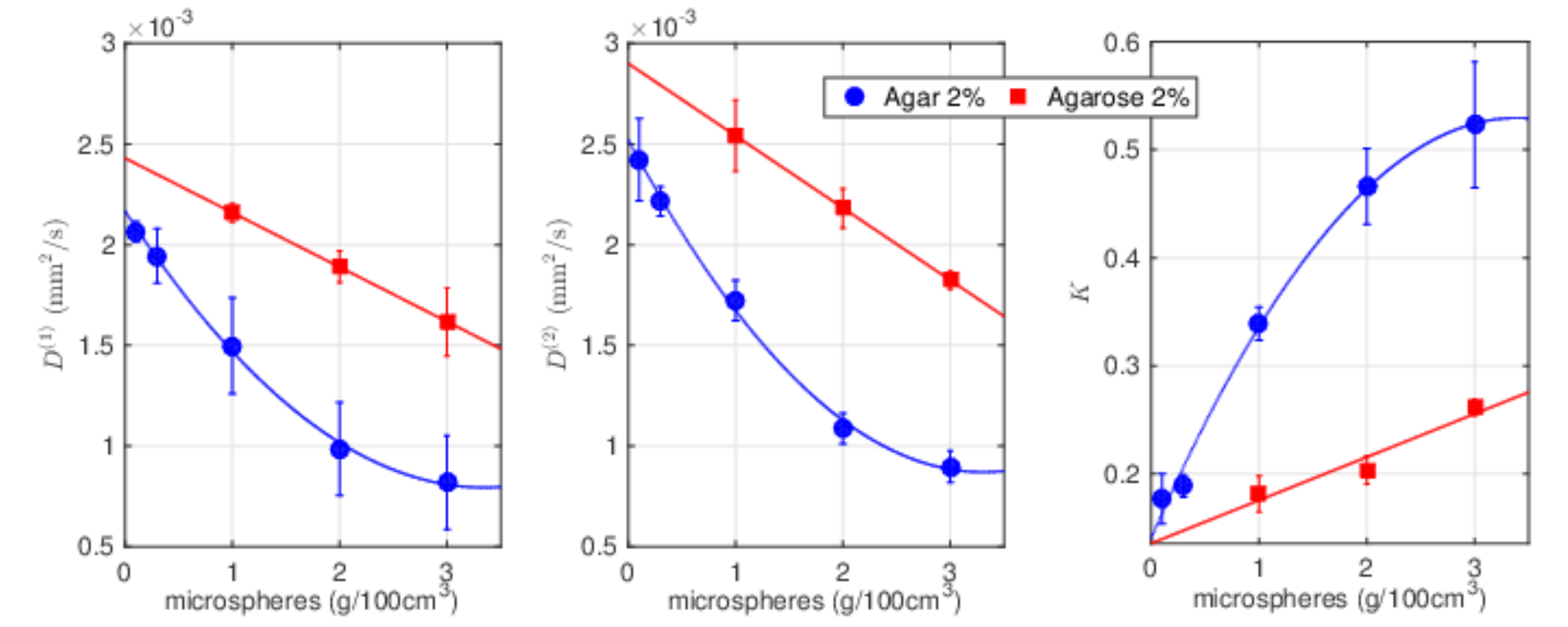}
\caption{Diffusion coefficients $D^{(1)}$ and $D^{(2)}$ and kurtosis $K$ for 2\% agar (blue dots) and 2\% agarose gels (red squares) as a function of microsphere concentration, with markers and error bars indicating mean values and 95\% confidence intervals obtained from the respective model fits for each gel, and dotted lines indicating general trends determined by linear (agarose) or quadratic (agar) fits of the diffusion coefficients and kurtosis as a function of microsphere concentration.} \label{fig:diffusion-vs-bead-conc}
\includegraphics[width=0.85\textwidth]{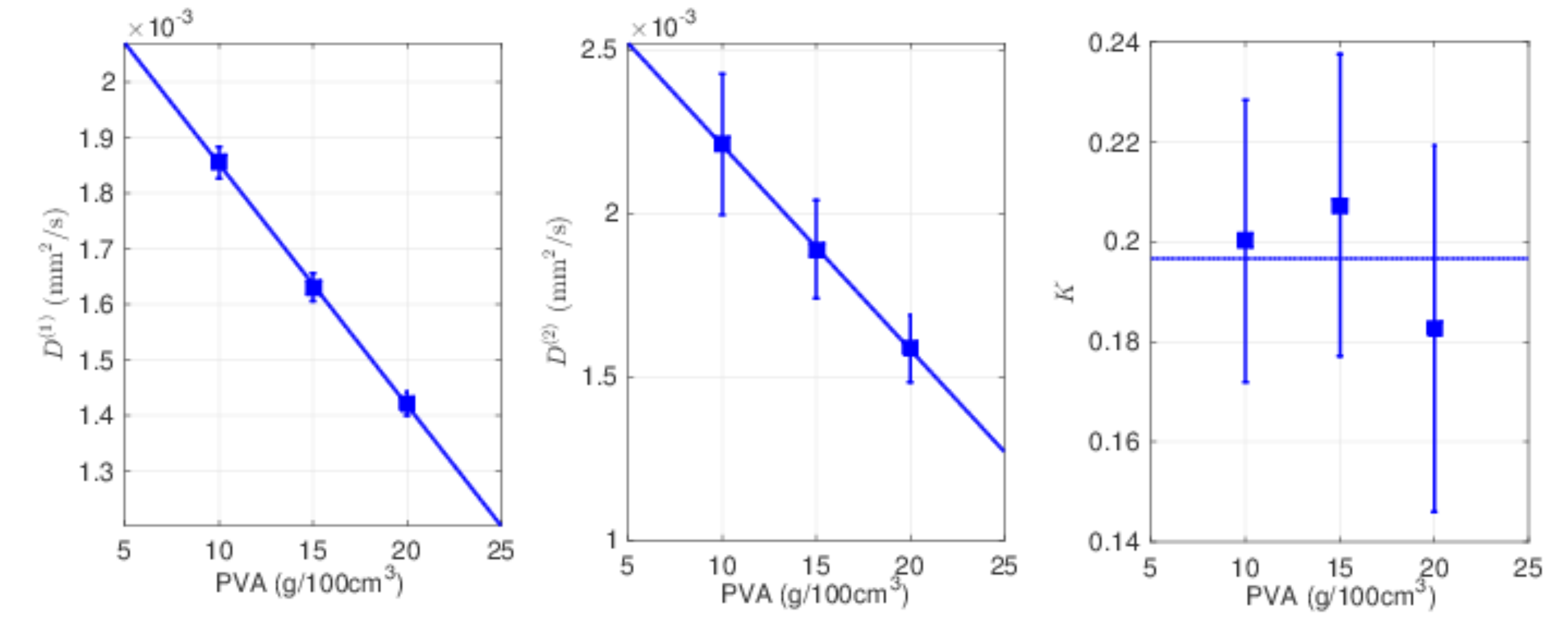}
\caption{Diffusion coefficients $D^{(1)}$ and $D^{(2)}$ and kurtosis $K$ as a function of the concentration of PVA with markers and error bars indicating mean values and 95\% confidence intervals obtained from the respective model fits for various pure PVA gels, and dotted lines indicating trends determined by linear regression fits of the diffusion coefficients and kurtosis as a function of PVA concentration.}
\label{fig:diffusion-PVA}
\end{figure*}

Fig.~\ref{fig:diffusion-vs-gel-conc} shows that the diffusion coefficients $D^{(1)}$ and $D^{(2)}$ tend to decrease or remain constant, while the kurtosis increases with the concentration of the gelling agent by approximately $0.035$ per gram of gelling agent per \SI{100}{cm^3} for pure agar and agarose gels, $0.040$ per gram of gelling agent per \SI{100}{cm^3} for agarose gels with \SI{1}{g} of microspheres added and $0.049$ per gram of gelling agent per \SI{100}{cm^3} for agar gels with \SI{2}{g} of microspheres added.  However, even for high concentrations the kurtosis of pure gels is limited, and well below the range of values observed for biological tissue shown in Table~\ref{table:kurtosis}.  For example, for pure agar gels the kurtosis obtained ranges from almost zero, $0.050(0.029,0.071)$, for a 1\% agar gel to only $0.126(0.100,0.152)$ for a 3\% agar gel, and similarly for pure agarose gels (see \SupTableII). 

The addition of \SI{2}{g} of microspheres per \SI{100}{cm^3} increases the kurtosis to $0.435 (0.415,0.456)$ for a 1\% agar gel and $0.523(0.490,0.556)$ for a 3\% gel.  For agarose gels the addition of \SI{1}{g} of microspheres increases the kurtosis from $0.196(0.183,0.209)$ for a 1\% agarose gel to $0.289(0.248,0.330)$ for a 3\% agarose gel.  \SupTableII\ further shows that even the addition of only \SI{0.1}{g} of microspheres per \SI{100}{cm^3} increases the kurtosis from $0.094(0.076,0.111)$ to $0.177(0.154,0.200)$ for a 2\% gel.  Similar increases are observed for other gels.

Fig.~\ref{fig:diffusion-vs-bead-conc} shows that the kurtosis can be varied by adjusting the concentration of the glass microspheres.  The variation is particularly large for the 2\% agar gels, increasing from $0.094(0.076,0.111)$ for a pure gel, to $0.523(0.465,0.881)$ when \SI{3}{g} of microbeads are added.  While the graph suggests a linear increase of the kurtosis with the concentration of microspheres for the 2\% agarose phantoms, the dependence of for the 2\% agar phantom is more complicated, characterized by a steeper initial increase in the kurtosis, followed by a levelling off at $\approx 0.52$ for microsphere concentrations of \SI{3}{g}.

For PVA phantoms Fig.~\ref{fig:diffusion-PVA} indicates a decrease in both $D^{(1)}$ and $D^{(2)}$ with the concentration of PVA similar to what is observed for the agar and agarose gels, however unlike for the former, the kurtosis does not appear to increase with concentation, fluctuating around $0.2$ with 95\% CIs ranging from $(0.172,0.228)$ for 10\% PVA to $(0.146,0.219)$ for 20\% PVA.

Spatially resolved diffusion and kurtosis maps derived from voxel-based analysis of the data, shown for a 2\% agar gel with \SI{2}{g} of glass microspheres in \SupFig6, indicate some spatial variation but still good homogeneity even for high concentrations of microspheres, as evidenced by the relatively narrow Gaussian distributions of the associated histograms for the diffusion and kurtosis parameters.  Specifically, we obtain 
  $D^{(1)} = (1459\pm 100) \times 10^{-6} \si{mm^2/s}$, i.e., $\Delta D^{(1)}/D^{(1)}\approx 6.8\%$,
  $D^{(2)} = (1669\pm 133) \times 10^{-6} \si{mm^2/s}$, i.e., $\Delta D^{(2)}/D^{(2)}\approx 8.0\%$, and 
  $K = 0.34\pm 0.02$, i.e., $\Delta K/K \approx 5.9\%$.

To assess the reproducibility of the results, multiple gels with the same composition were produced in a few cases.  Comparison of the results for two 2\% agarose gels with \SI{2}{g} microspheres in \SupFig8 shows that the diffusion coefficients $D^{(1)}$ differ by 5\%, the diffusion coefficients $D^{(2)}$ by about 3\%, and the kurtosis values by about 10\%.  Thus, there is some variability but the results are reproducible to within a few percent, and could probably be further improved by enhanced process control of the preparation of the gels.

\subsection{Relaxation Properties}

\begin{figure*}
\centering
\includegraphics[width=0.80\textwidth]{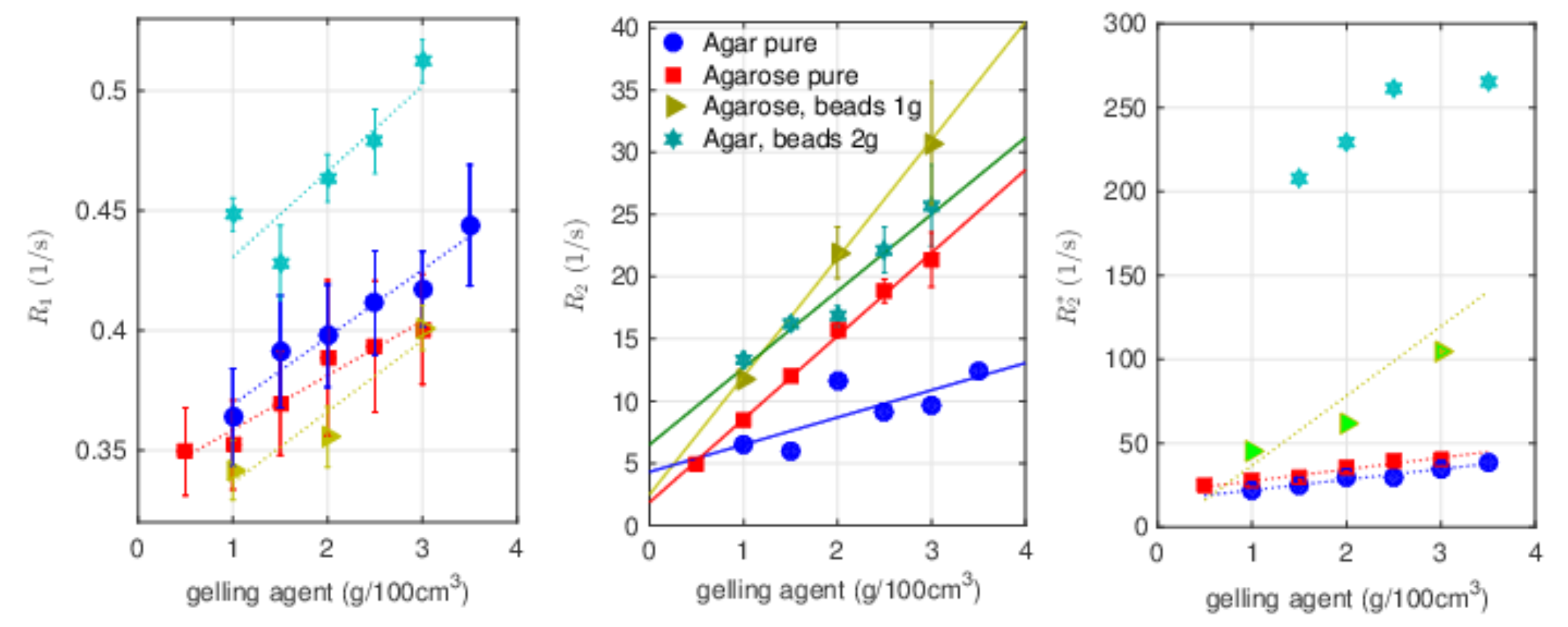}
\caption{$R_{1}$, $R_{2}$ and $R_2^*$ as a function of the gelling agent concentration, with markers and error bars indicating mean values and 95\% confidence intervals obtained from the respective model fits for various pure agar and agarose gels (blue dots and red squares, respectively) as well as agar and agarose gels with microspheres (green stars and yellow triangles, respectively).  The dotted lines indicate trends determined by linear regression fits of the relaxation rates as a function of gelling agent concentration.} \label{fig:R-vs-gel-conc}

\includegraphics[width=0.80\textwidth]{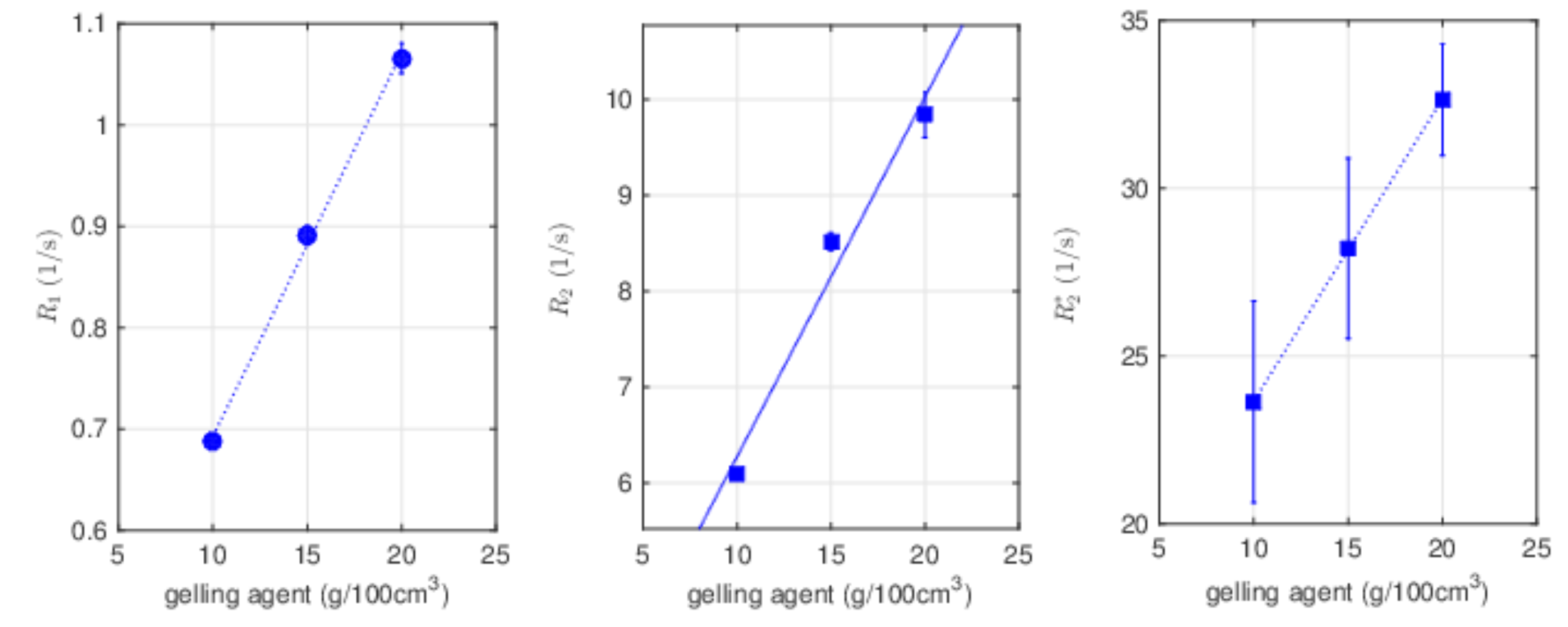}
\caption{$R_{1}$, $R_{2}$ and $R_2^*$ as a function of the concentration of PVA with markers and error bars indicating mean values and 95\% confidence intervals obtained from the respective model fits for various pure PVA gels, and dotted lines indicating trends determined by linear regression fits of the relaxation rates as a function of PVA concentration.}\label{fig:R-PVA}

\includegraphics[width=0.80\textwidth]{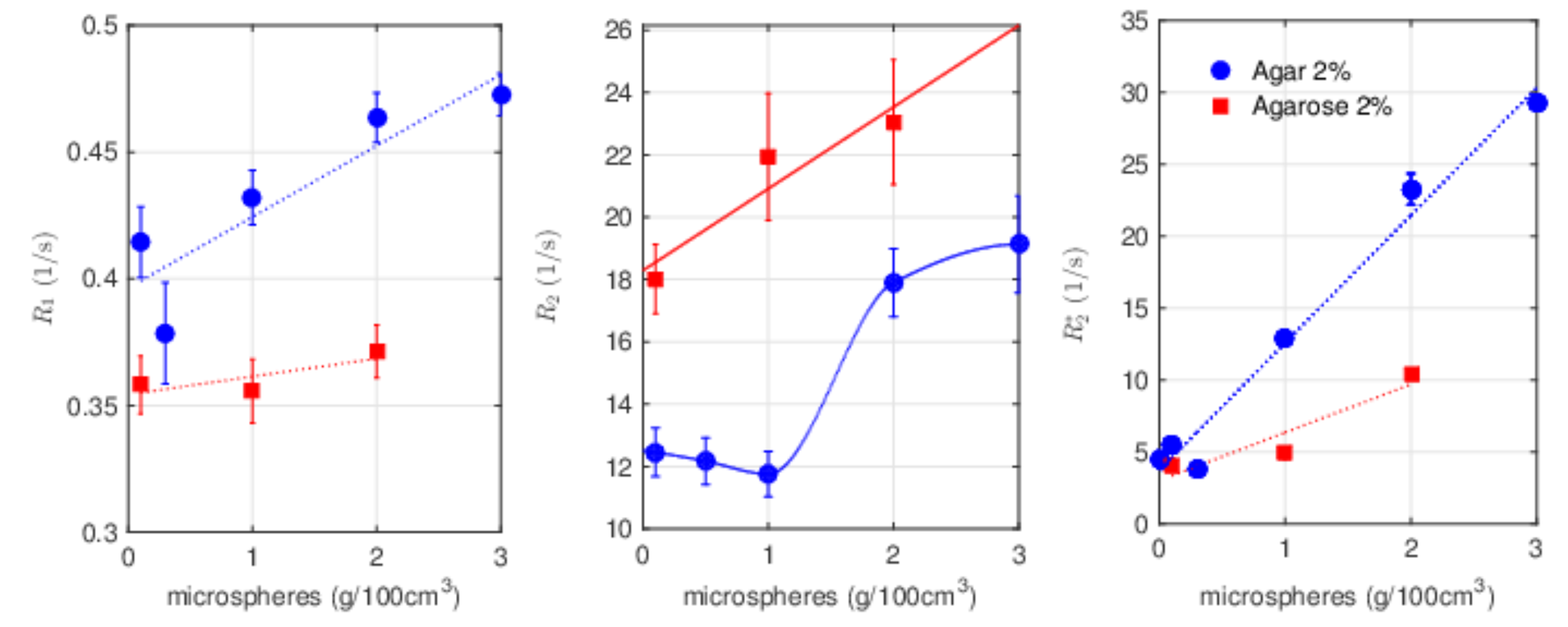}
\caption{$R_{1}$, $R_{2}$ and $R_2^*$ for 2\% agar (blue dots) and 2\% agarose gels (red squares) as a function of microsphere concentration, with markers and error bars indicating mean values and 95\% confidence intervals obtained from the respective model fits for each gel, and dotted lines indicating general trends determined by fits of the relaxation rates as a function of microsphere concentration.}\label{fig:R-vs-bead-conc}
\end{figure*}

For pure gels $R_{1}$ was found to range from $0.35 (0.33,0.37)$ to $0.44 (0.42,0.47)$~\si{1/s}, $R_2$ from $4.9(4.63,5.16)$ to $22.14(19.9,24.39)$~\si{1/s}, and $R_2^*$ from $21.89(20.69,23.1)$ to $40.58(38.29,47.88)$~\si{1/s}.  For gels with microspheres, $R_{1}$ ranged from $0.34 (0.32,0.35)$ to $0.51 (0.50,0.52)$~\si{1/s}, $R_2$ from  $9.69 ( 9.34,10.04)$ to $33.47(27.82,39.13)$~\si{1/s}, and $R_2^*$ from $21.89 (20.69,23.10)$ to $246.50(232.50,260.40)$~\si{1/s}.

Figures \ref{fig:R-vs-gel-conc} and \ref{fig:R-PVA} show that $R_{1}$, $R_2$ and $R_2^*$ increase linearly with the concentration of the gelling agent for all gels (agar, agarose, PVA), as expected.  However, $R_{1}$, $R_2$ and $R_2^*$ also have a dependence on the concentration of microspheres.  Fig.~\ref{fig:R-vs-bead-conc} shows that for a fixed concentration of the gelling agent, $R_{1}$, $R_2$ and $R_2^*$ exhibit approximately linear increases with microsphere concentration with the notable exception of $R_2$ for the 2\% agar gels, which exhibits an anomalous dip around \SI{1}{g}.  For example, for an agar concentration of 2\%, $R_{1}$ increases by \SI{0.03}{1/s} per \SI{1}{g} of microspheres added, $R_2$ increases by \SI{1.49}{1/s} and $R_2^*$ by \SI{87.357}{1/s}.  This effect is also observed in agarose, but to a lesser extent. Full details about the  $R_1$, $R_2$ and $R_2^*$ values obtained for all phantoms, together with statistical information (95\% confidence intervals for each parameter) are given in \SupTableIII.

\section{Discussion}

The results show that there is evidence of non-Gaussian diffusion in all of the gel phantoms, consistent with the idea that macromolecular structures formed by the gelling agents constitute barriers to the diffusive motion of the water molecules.  The observed kurtosis values also  increase with the concentration of gelling agent, at least for agar and agarose gels, but the kurtosis values for the pure gel phantoms are too low to mimic the values observed in the literature \textit{in vivo} for a variety of human tissues.

The results further show that the addition of glass microspheres to the gels can considerably increase the observed kurtosis values, especially for agar gels, resulting in phantoms with kurtosis values matching those found \textit{in vivo} more closely.  This is consistent with the expectation that glass microspheres increase the barrier concentration and observations from other studies that increased barrier concentration increases kurtosis~\cite{Novikov2011,Chu-Lee2013}.

The effect is more pronounced for agar gels.  This may be due to the fact that agarose, one of the two principal components of agar,  purified from agar by removing agar's other component, agaropectin, forms thermo-reversible gels consisting of thick bundles of agarose chains linked by hydrogen bonds, with large pores holding water.  Water in the large pores tends to relax more slowly than water in small pores because of the different relative amounts surface and bulk water. It has also been noted that diffusion of particles in agarose gels is anomalous, with a diverging fractal dimension of diffusion when large particles become entrapped in the pores of the gel~\cite{Narayanan2006,Fatin-Rouge2004,Ioannidis2009}.

Characterization of the gel phantoms using relaxation rate measurements to quantify the effect of the glass microspheres on $R_{1}$, $R_2$ and $R_2^*$ of the phantoms show that the addition of glass microspheres increases the observed relaxation rates, but the values remain suitable for tissue-mimicking phantoms, with the relaxation rates for a 1\% agar gel with \SI{2}{g} of microspheres being similar to those of 3.5\% agar gel.  Thus the relaxation rates can be controlled by adjusting the concentration of the gelling agent and $R_1$ modifiers could be added if necessary.

With regard to spatial homogeneity, voxel-based analysis of the relaxation and diffusion data indicates that the spatial variation of  the diffusion parameters, as quantified by the standard deviation of the parameter over its mean value, is on the order of a few percent in the examples studied.  Characterization of the microstructure of the gel using high-resolution micro-CT or optical microscopy would be an interesting avenue for future research. 

Although more extensive, systematic longitudinal studies would be desirable, repeat scans performed after more than six months of storage of the gels at room temperature suggest good long-term stability of both the gels, and their relaxation and diffusion properties (see Table~\ref{table:3} and \SupTableIII). This is consistent with the observations in the literature regarding the geometric stability of agar phantoms~\cite{Madsen2005}.

Other limitations of the current study are that the data was acquired on a single scanner with a particular combination of coils, and the kurtosis values may be biased towards higher values due to low SNR for high $b$ values, compounded by limited bit depth resolution of the DICOM images used for the quantitative analysis.  To improve the accuracy of quantitative phantom parameters cross-platform scan protocols for multi-site studies should be developed, and more work is needed on improving SNR in DWI for high $b$-values and characterization of noise-induced bias.  Finally, it would be desirable to investigate the effect of pulse sequence parameters such as the duration and spacing of the diffusion pulses on the kurtosis results.

\FloatBarrier
\section{Conclusions}

We investigated the diffusion properties of gel phantoms commonly used in MRI including agar, agarose and PVA gels with emphasis on the characterization of non-Gaussian diffusion as quantified by the kurtosis.  While pure gels were found to have low kurtosis values, we demonstrated that the kurtosis can be increased considerably by the addition of glass microspheres and controlled by varying the microsphere concentration without substantial changes to the consistency, homogeneity or relaxation properties of the phantoms.

The work suggests that agar gel phantoms with glass microspheres in particular are promising materials for inexpensive but durable gel phantoms for DKI with tuneable diffusion and relaxation properties. Future work includes the study of the effects of different types of glass microspheres and the design of more complex structured phantoms with non-isotropic, non-Gaussian diffusion for multi-modal imaging.  Nuclear Magnetic Resonance (NMR) and multi-center studies of novel materials for DKI phantoms will be required to assess inter-platform variability and establish reliable material characteristics, optimal protocols and develop improved EPI-based sequences.


\begin{acknowledgments}
We gratefully acknowledge useful conversations with Geraint Lewis.  ZGP acknowledges TUBITAK ``The Scientific and Technological Research Council of Turkey'' Project No. 1059B141400677 and Research Fund of Cukurova University, Project No. FDK-2014-2709 for financial support. SS acknowledges funding from a Royal Society Leverhulme Trust Senior Fellowship.  
\end{acknowledgments}

\bibliographystyle{ama}


\cleardoublepage

\setcounter{section}{0}
\setcounter{figure}{0}
\setcounter{table}{0}
\setcounter{page}{1}
\section{Supplementary Material}

\subsection{Comparison of Diffusion Sequences for Water Phantom}
\label{appendix:water-control}

\begin{figure*}[!h]
  \subfloat[Vendor EPI diffusion sequence]  {\includegraphics[width=\textwidth]{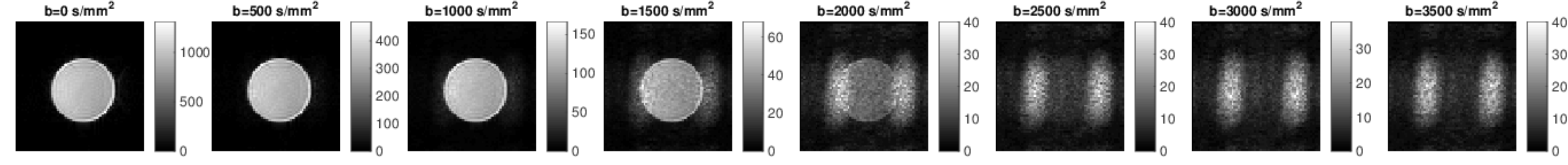}}\\
  \subfloat[Custom non-EPI PGSE sequence]{\includegraphics[width=\textwidth]{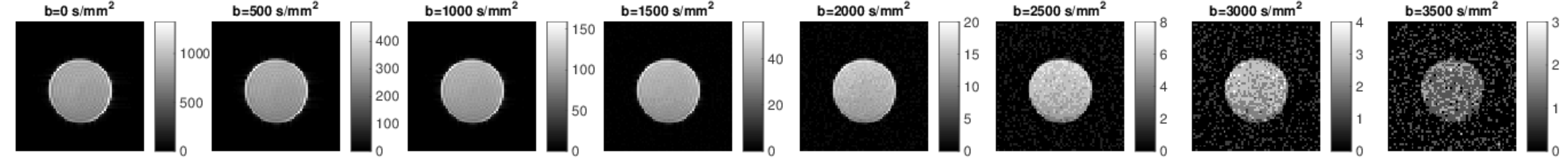}}
  \caption{Comparison of vendor-supplied and custom diffusion sequences for vendor-supplied water phantom.  The images generated with the vendor-supplied EPI diffusion sequence suffer from large artifacts and almost complete signal loss for $b$-values as low as \SI{2500}{s/mm^2}.  For our custom sequence no such artifacts are observed and the phantom is still clearly visible for much higher $b$-values.}
 \label{fig:EPIvsPGSE}

 \subfloat[Vendor EPI sequence]    {\includegraphics[width=0.85\columnwidth]{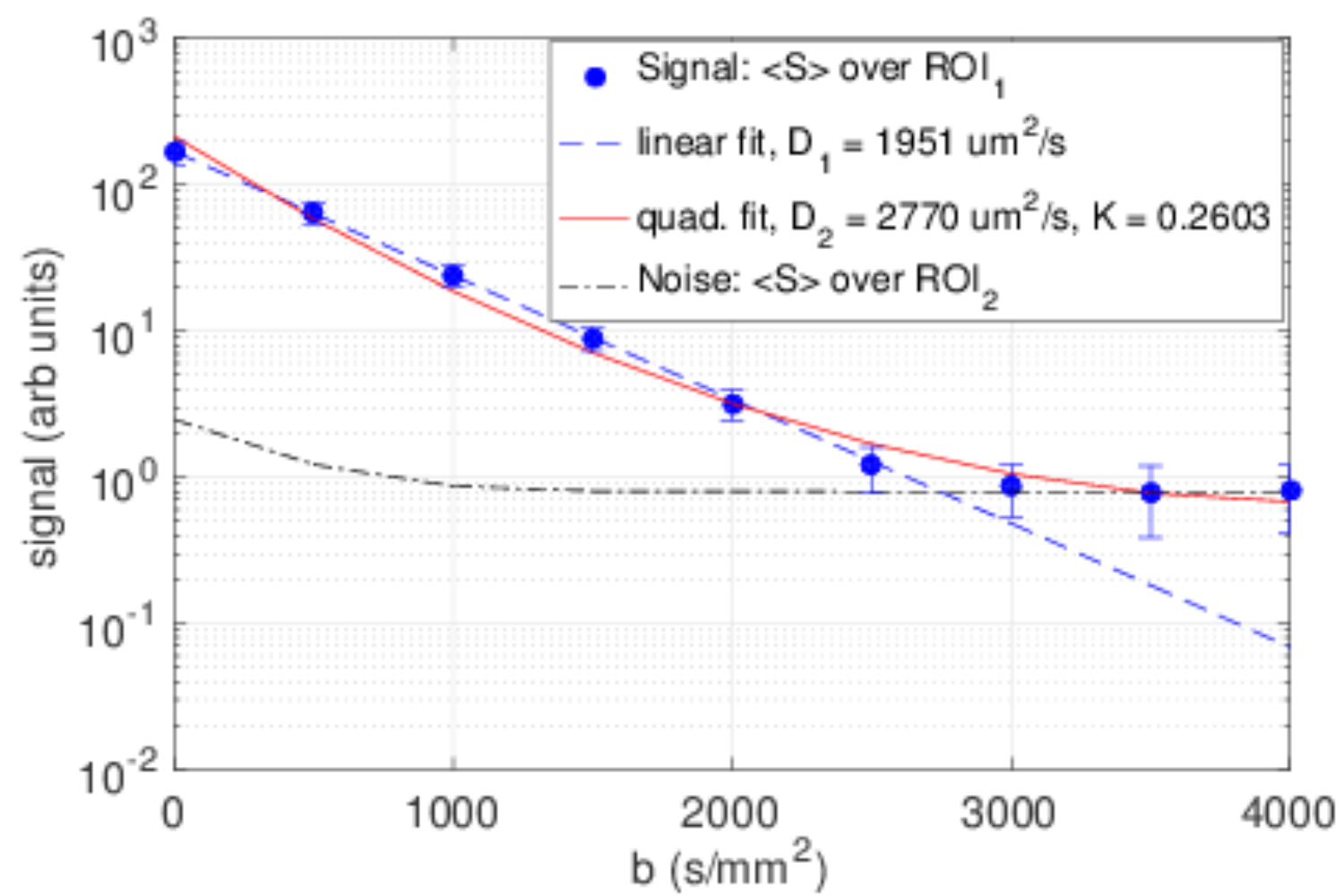}} \qquad
 \subfloat[Custom PGSE sequence]{\includegraphics[width=0.85\columnwidth]{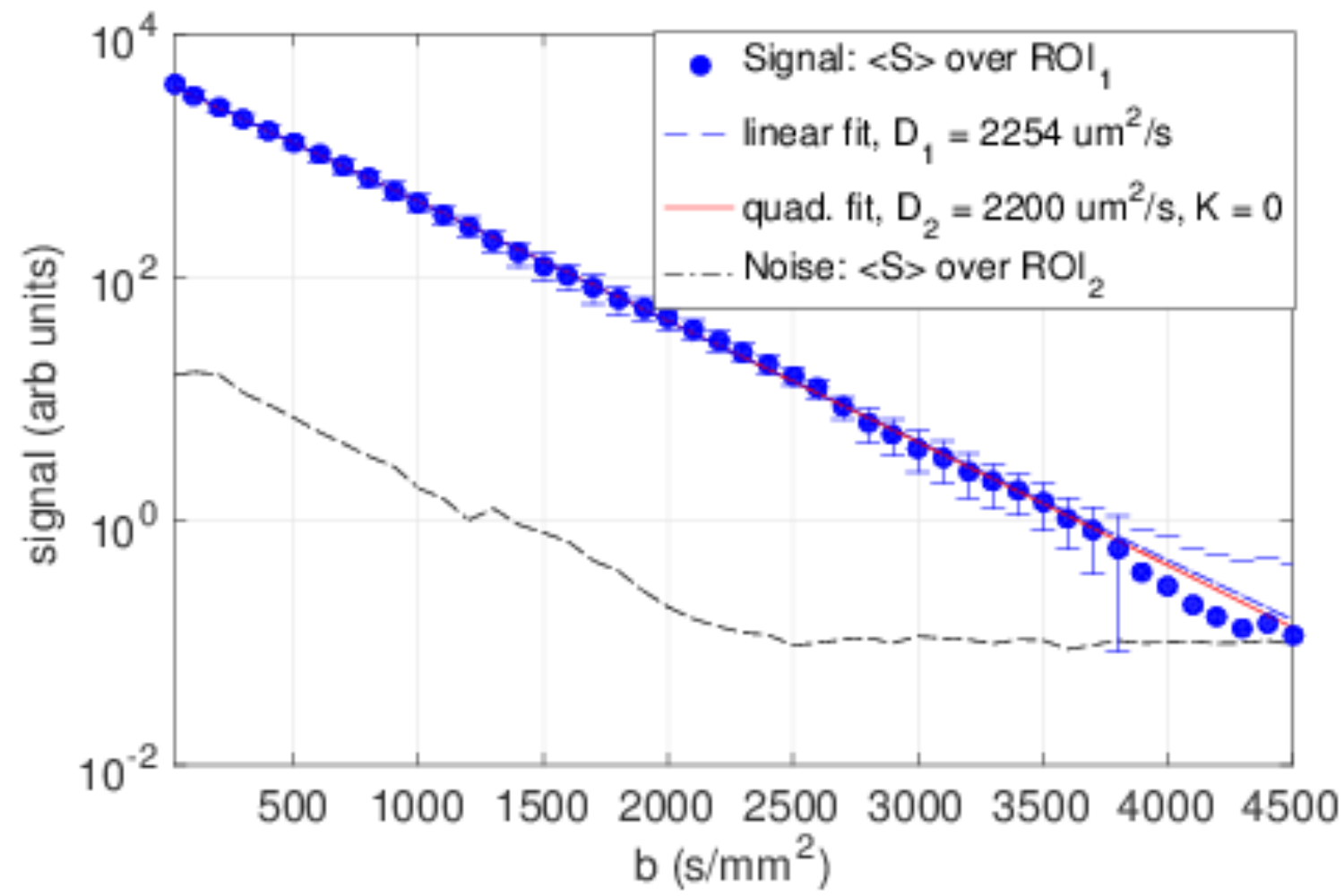}}
 \caption{The mean signal for vendor supplied water phantom as a function of the strength of the diffusion gradients shows vanishing phantom-air contrast for $b$-values as low as \SI{2500}{s/mm^2} for the vendor sequence, consistent with Fig.~\ref{fig:EPIvsPGSE}, which can lead to pseudo-kurtosis unless care is taken to exclude these $b$-values from the fit.  For our custom sequence good phantom-air contrast is maintained for $b$-values up at least \SI{3500}{s/mm^2} and the diffusion coefficients $D^{(1)}$ and $D^{(2)}$ and kurtosis obtained are in excellent agreement with the literature values.} \label{fig:water}

  \subfloat[Histograms of pixel values for ROI1]{\includegraphics[width=0.85\columnwidth]{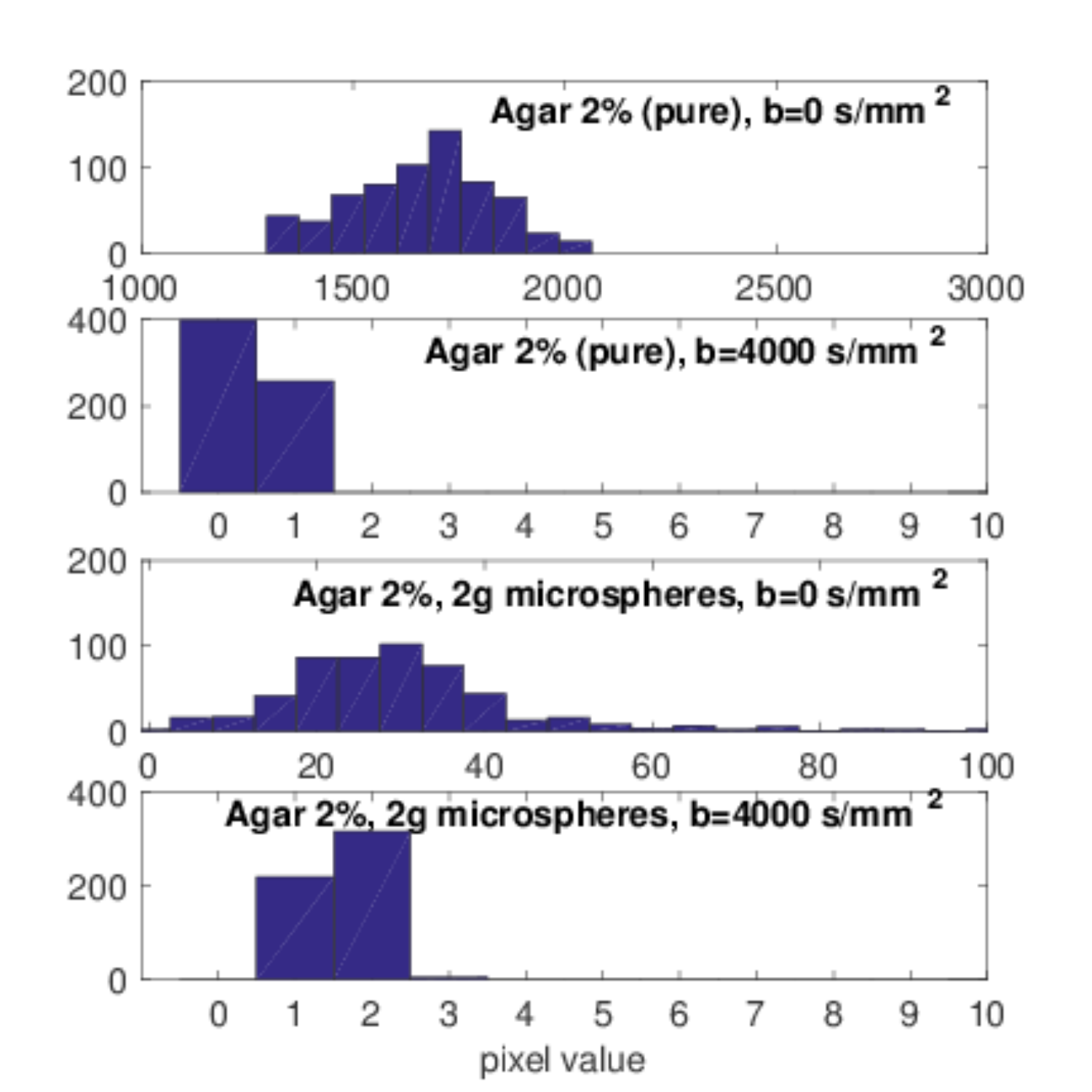}} \qquad
  \subfloat[Histograms of pixel values for ROI2]{\includegraphics[width=0.85\columnwidth]{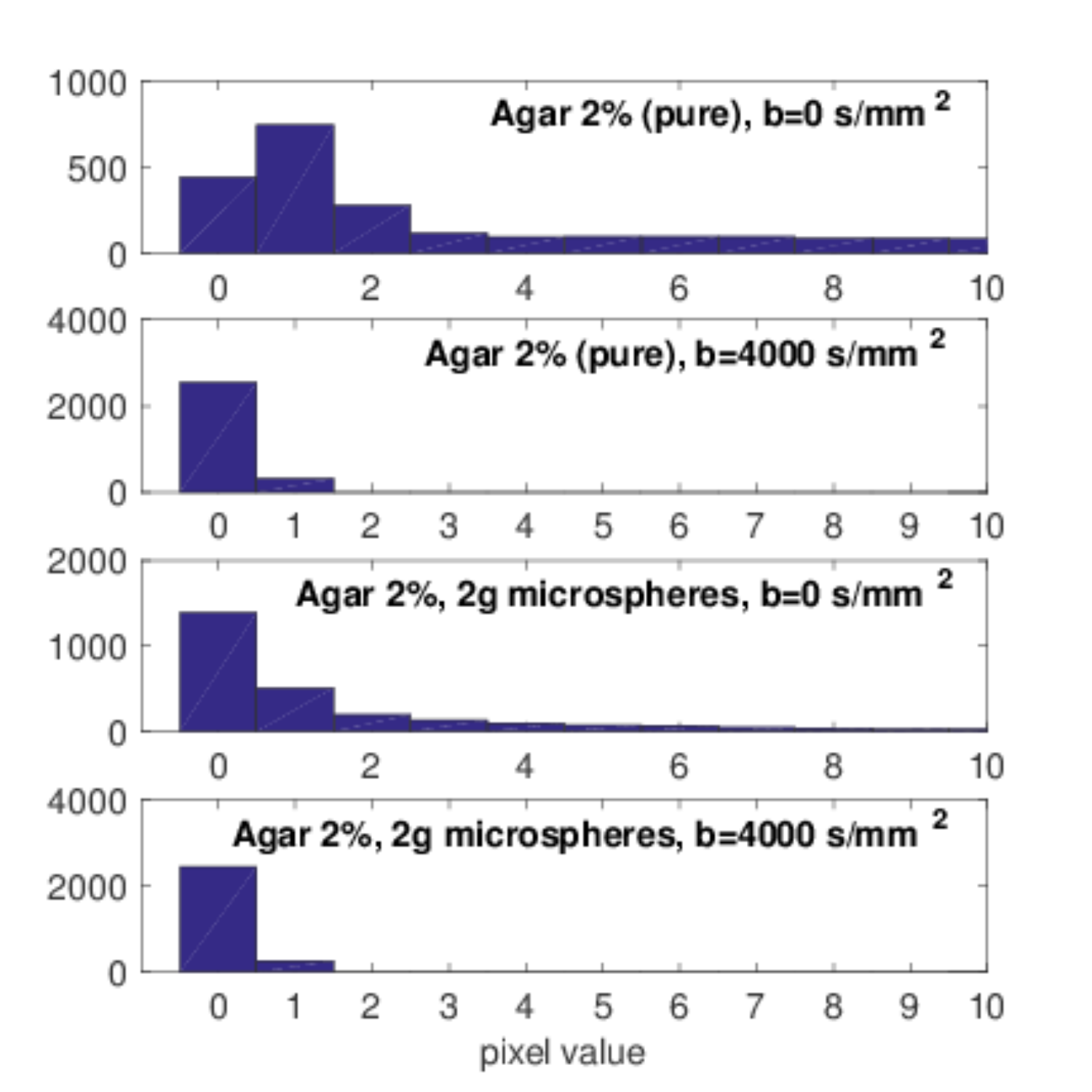}}
  \caption{Distribution of pixel values for phantom (ROI1) and air (ROI2) regions for two agar gels for $b=\SI{0}{s/mm^2}$ and $b=\SI{4000}{s/mm^2}$, showing that the distributions for the noise ROI and high $b$-values deviate from the theoretically expected Rician distribution.}
  \label{fig:histograms}
\end{figure*}

Our custom PGSE sequence was tested and compared to the vendor-supplied diffusion sequence for a water phantom as a negative control.  The common scan parameters for both sequences were: Field-of-view (FOV) \SI{100}{mm} $\times$ \SI{100}{mm}, matrix size $64\times 64$, $T_{E}= \SI{120}{ms}$, single average.  For our custom PGSE, $T_{R}=\SI{3000}{ms}$, a readout bandwidth of \SI{130}{Hz/pixel} and a slice thickness of \SI{20}{mm} were chosen.  For the vendor-supplied EPI diffusion sequence, the maximum slice thickness of \SI{10}{mm} and $T_R=\SI{5000}{ms}$ were selected and various combinations of readout bandwidths and phase encoding steps and other parameters were tested.  The images in Fig.~\ref{fig:EPIvsPGSE}(a) were acquired with  $56$ phase encoding steps with an echo train length of $28$ and a readout bandwidth of \SI{1200}{Hz/px}.

Fig.~\ref{fig:EPIvsPGSE}(a) shows that images obtained with the standard vendor-supplied EPI diffusion sequence suffer from large artifacts, especially in the phase encode direction, for $b$ values as low as \SI{1500}{s/mm^2} and complete signal loss for $b$ values $\ge \SI{2500}{s/mm^2}$.  Fig.~\ref{fig:EPIvsPGSE}(b) shows  that for our custom PGSE no artifacts are visible and phantom-air contrast is maintained for $b$-values up to at least \SI{3500}{s/mm^2}.

Fig.~\ref{fig:water} further shows that the mean signal for a water phantom effectively vanishes for $b>$\SI{2500}{s/mm^2} for the vendor-supplied EPI sequence, while the signal for our custom PGSE remains well above the background noise level for $b$-values up to \SI{4000}{s/mm^2}.  For $b$ values up to \SI{2000}{s/mm^2} we obtain a good linear fit in both cases but the ADC obtained is closer to the literature value of \SI{2200}{mm^2/s} for water for our custom PGSE sequence.  For higher $b$-values ``pseudo-kurtosis'' may be observed for the EPI-based product sequence, consistent with previous observations \cite{Scheel2015,Portakal2017}, while no such effect is observed for our custom PGSE sequence, which indicates effectively no kurtosis even for high $b$-values, as expected for a water phantom.

\begin{figure} \!\!\!\!\!\!
  \subfloat[CDF of pixel values for ROI1]{\includegraphics[width=0.525\columnwidth]{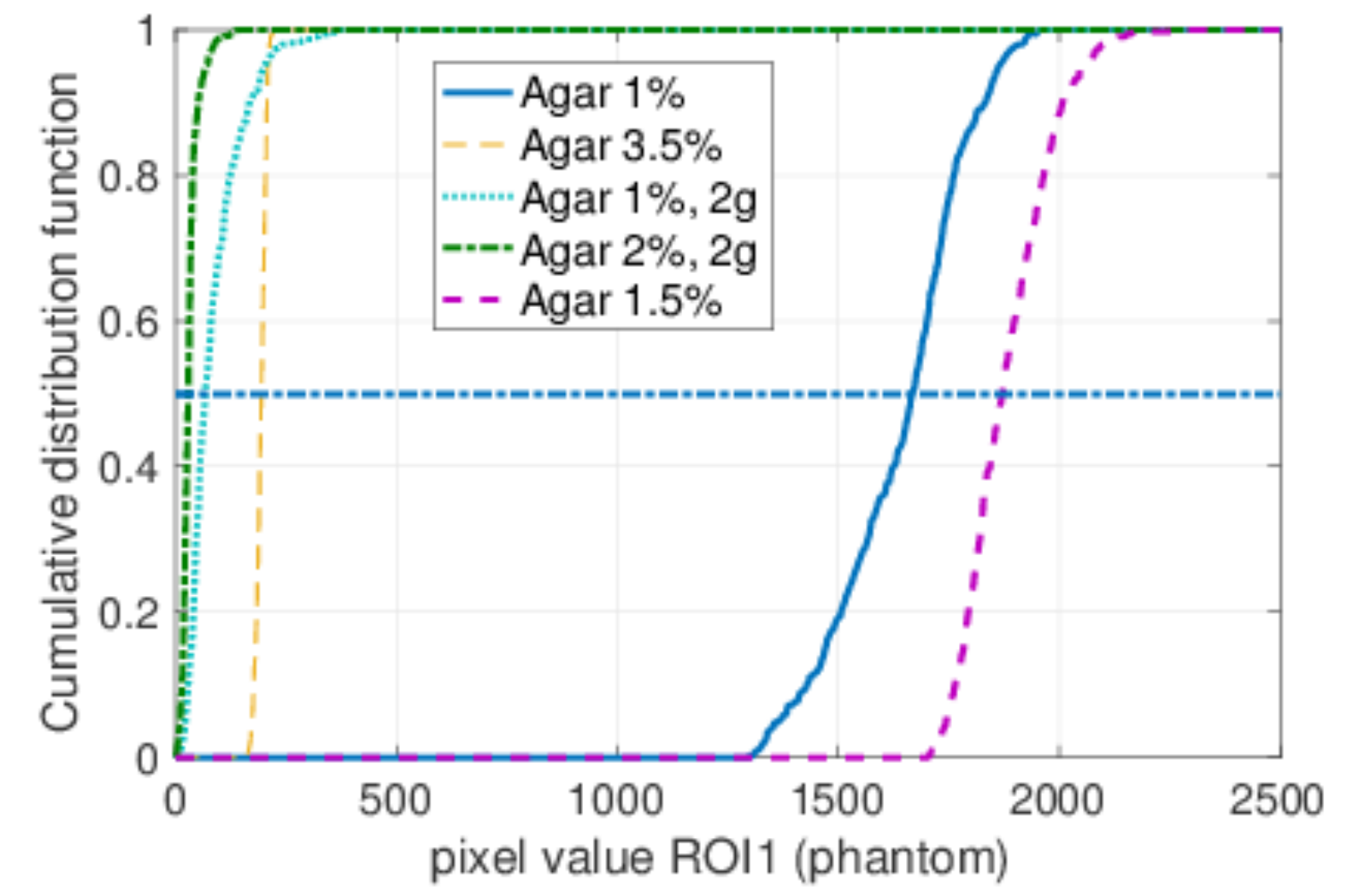}} \!\!
  \subfloat[CDF of pixel values for ROI2]{\includegraphics[width=0.525\columnwidth]{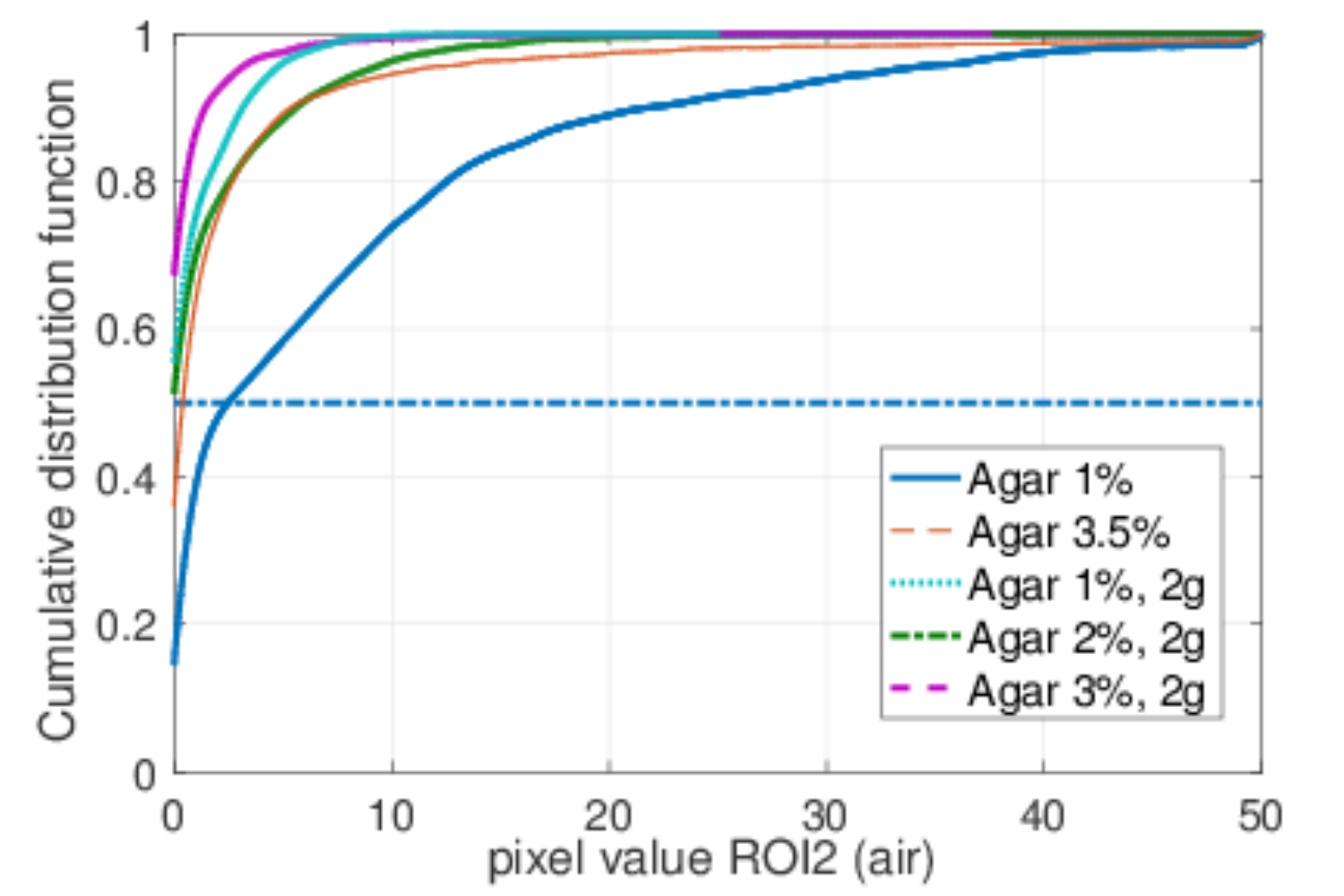}} \!\!\!\!\!\!\!\!
  \caption{Cumulative distribution function (CDF) for various agar gels for phantom (ROI1) and air (ROI2) regions, used to calculate
    phantom-air contrast, for $b=\SI{0}{s/mm^2}$.}\label{fig:CPD}
\end{figure}

\subsection{Detailed Noise Analysis}
\label{appendix:noise-quantification}

To avoid overestimation of kurtosis it is important to avoid noise floor fitting.   The noise level is usually characterized by computing the signal-to-noise ratio (SNR).  NEMA guidelines~\cite{NEMA} provide four different methods to estimate SNR.  The preferred choice is method 1, which requires the acquistion of two images with the same acquisition parameters and computation of a difference image.  The SNR is then defined as $S/\sigma$ where $S$ is the mean signal over the ROI of the first image, and $\sigma$ is the standard deviation of the difference image over the ROI divided by $\sqrt{2}$.  We calculated the SNR for a series of DWI images with different $b$-values for two of the gels with the highest observed kurtosis using this method.  For comparison the same SNR calculations were performed using both the DICOM images produced by the standard image reconstruction software on our scanner, as well as images manually reconstructed from the raw data acquired.  The results, shown in Table~\ref{table:SNR}, suggest that, unlike for the EPI sequence, there is still sufficient SNR even for the highest $b$-values used.  It is interesting to note that the SNR calculated from the standard DICOM images is lower than the SNR obtained from images reconstructed from the raw data.  This appears to be due to the fact that the default settings for the image reconstruction software on our scanner limit the bit depth to 12 bits.  However, processing of raw data is usually time-consuming and not standard in clinical practice.  Accordingly, the results presented in the main part of the paper were based on the standard DICOM images produced by the scanner software.

\begin{table}
  \begin{tabular}{|l||*{9}{c|}} \hline 
    $b$ (\si{s/mm^2}) & 0     &  500 & 1000 & 1500 & 2000 & 2500 & 3000 & 3500 & 4000 \\\hline 
    Gel 1 (D) & 30.8 &  46.7 &  30.8 &  19.9 &  14.6 &  11.2 &   8.9 &   6.5 & 5.0 \\\hline
    Gel 1 (R) & 31.8 &  54.5 &  37.3 &  25.8 &  19.4 &  15.3 & 12.6 & 10.1 & 8.0 \\\hline
    Gel 2 (D) & 23.1 &  14.4 &  10.2 &    7.1 &    5.0 &    5.1 &   3.8 &   2.9 & 3.0 \\\hline
    Gel 2 (R) & 27.6 &  19.6 &  14.2 &  10.4 &    8.0 &    6.8 & 6.6   &   5.3 & 5.2\\\hline
  \end{tabular}
  \caption{SNR for two gel phantoms with high kurtosis calculated using NEMA method 1 from Siemens DICOM (D) images and raw data (R), respectively. Gel 1: Agar 2\%, \SI{2}{g} microspheres, Gel 2: Agar 3\%, \SI{2}{g} microspheres.}  \label{table:SNR}
\end{table}

It was also not practical to acquire duplicate images for every single gel and diffusion weighting.   Therefore NEMA method 4 was considered as an alternative to estimate SNR from a single image by selecting a phantom ROI and an air ROI for each image, and calculating the ratio of the mean signal over the phantom ROI and the standard deviation over the air ROI.  NEMA guidelines for method 4 stipulate that the phantom ROI should cover at least 75\% of the phantom and the ``air'' ROI should be as large a size as possible in the background, while avoiding areas corrupted by artifacts~\cite{NEMA}.   These conditions were generally satisfied by our automated ROI selection algorithm, which used thresholding to locate the sample and select a circular ROI about 80\% of the diameter of the sample (ROI1).  The ``air'' ROI (ROI2) was defined as all voxels outside a circle approximately $1.2$ times the diameter of the sample, as shown in Fig.~\ref{fig:images}.  As artifacts in the phase-encode direction are negligible for our custom PGSE sequence even for high $b$-values, this ensures that an ``air'' ROI as large as possible while avoiding artifacts.

However, while the histograms resemble a Gaussian (Rician) distribution and the mean of the signal over the phantom region (ROI1) is well defined for $b=\SI{0}{s/mm^2}$, Fig.~\ref{fig:histograms}(a) shows that for high $b$ values discretization effects due to the limited bit-depth of the standard DICOM images modify the observed distribution.  These effects are even more pronounced for the air ROI (Fig.~\ref{fig:histograms}(b)).  While the distribution for the 2\% plain agar at $b=\SI{0}{s/mm^2}$ still resembles the theoretically expected Rayleigh distribution for the air region, for denser gels or higher $b$-values this is not necessarily the case.  Simply computing the standard deviation of the distribution in this case is problematic.  The cumulative distribution functions, shown in Fig.~\ref{fig:CPD}, can be used to determine the threshold $\theta$ such that at least $x$\% of the signal values in the air region are below $\theta$.  Alternatively, motivated by our water phantom results, we use phantom-air contrast, which we define as mean signal over the phantom region (ROI1) divided by the mean signal over the air region (ROI2), as a practical measure to decide which $b$-values should be considered in the ADC/kurtosis fit from single DICOM images.

\subsection{Homogeneity, stability and reproducibility}
 
\begin{figure} \!\!\!\!\!\!\!
  \subfloat[$R_2^*$ vs slice position]{\includegraphics[width=0.53\columnwidth]{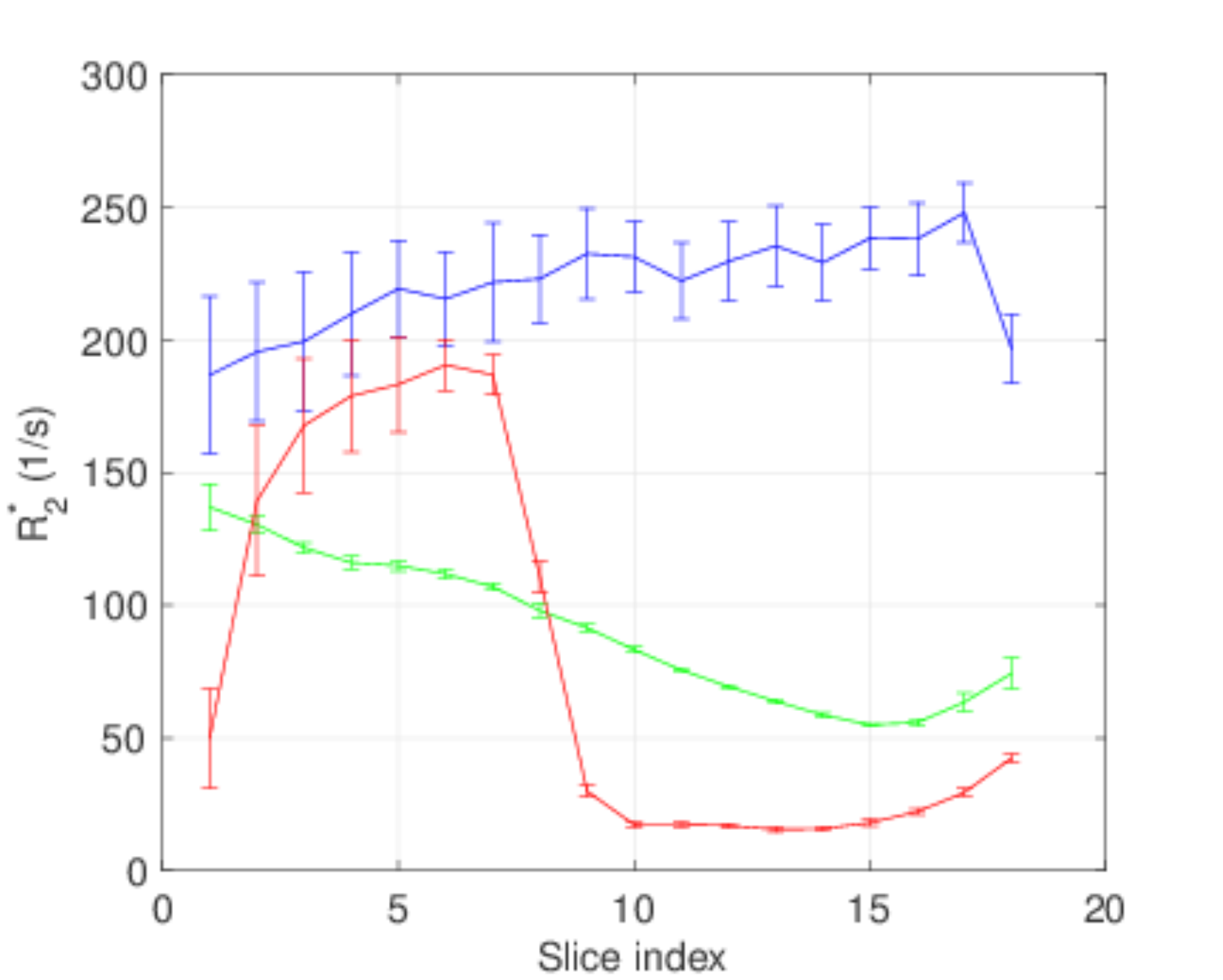}}\!\!\!\!\!\!\! 
  \subfloat[Double Echo Field Map]{\includegraphics[width=0.53\columnwidth]{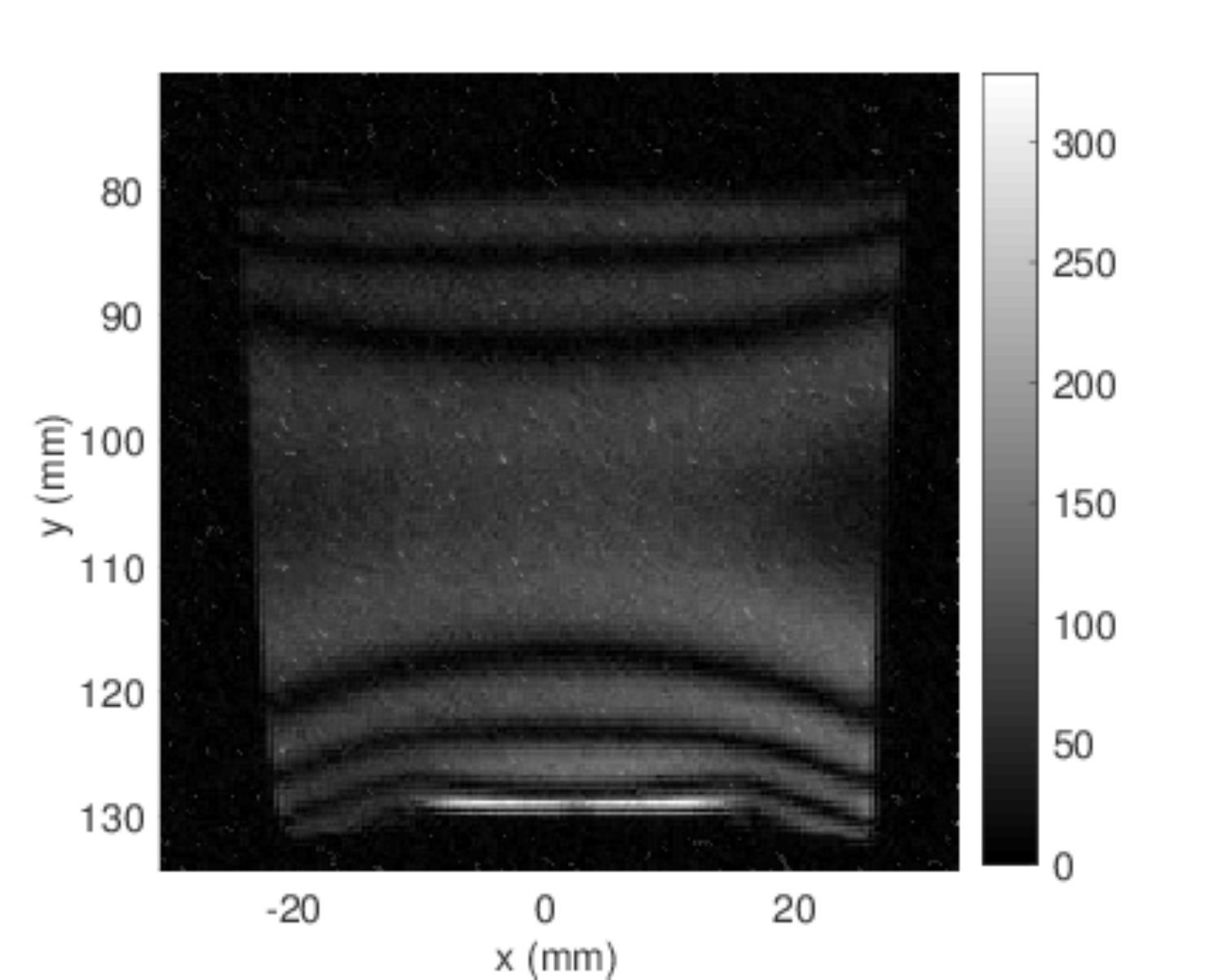}} \!\!\!\!\!\! 
  \caption{Variation of $R_2^*$ in the axial direction for three gels (a) and typical double echo interference field map for a
    transverse slice through center illustrating $B_0$ field inhomogeneity.  The distance between adjacent dark fringes in the
    field map corresponds a Lamor frequency difference between these points of \SI{50}{Hz}. }
    \label{fig:R2s+Fieldmap}
\end{figure}

\begin{figure*} 
  \subfloat[$D^{(1)}$ Map]{\scalebox{0.4}{\adjustbox{trim={0.05\width} {0.15\height} {0.05\width} {0.01\height},clip}{%
        \includegraphics{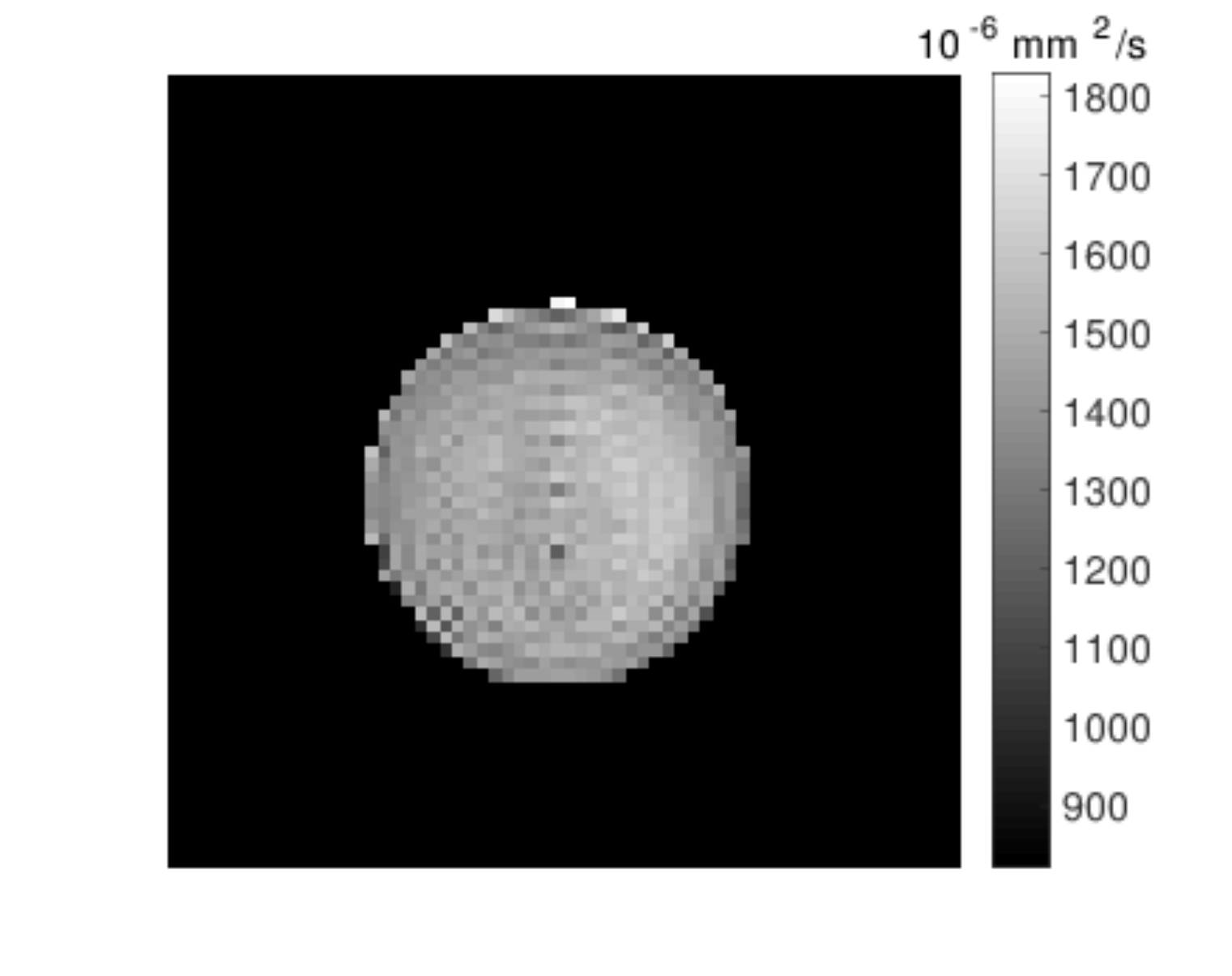}}}} 
   \subfloat[$D^{(2)}$ Map]{\scalebox{0.4}{\adjustbox{trim={0.05\width} {0.15\height} {0.05\width} {0.01\height},clip}{%
         \includegraphics{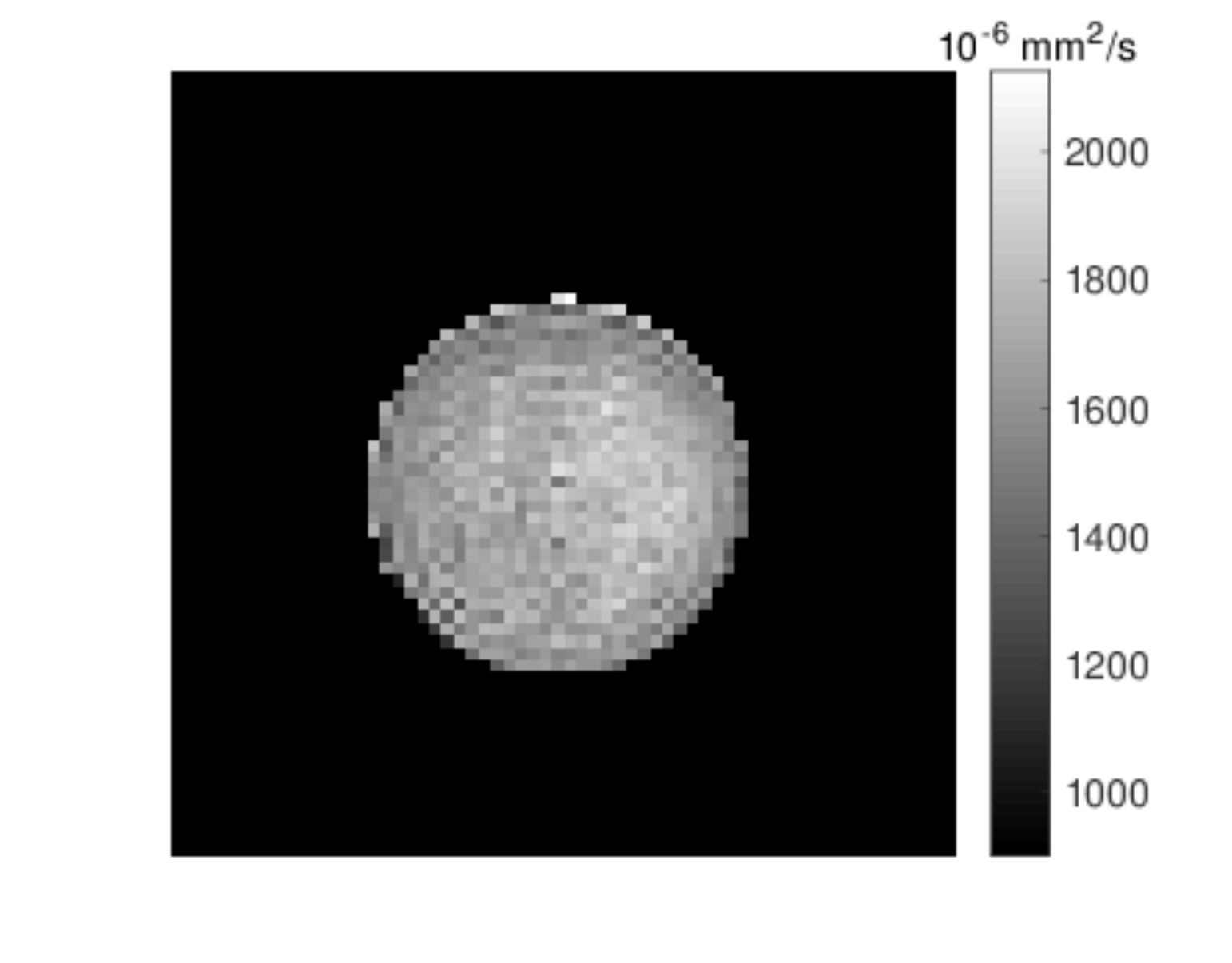}}}}
  \subfloat[Kurtosis Map]{\scalebox{0.4}{\adjustbox{trim={0.05\width} {0.15\height} {0.05\width} {0.01\height},clip}{%
        \includegraphics{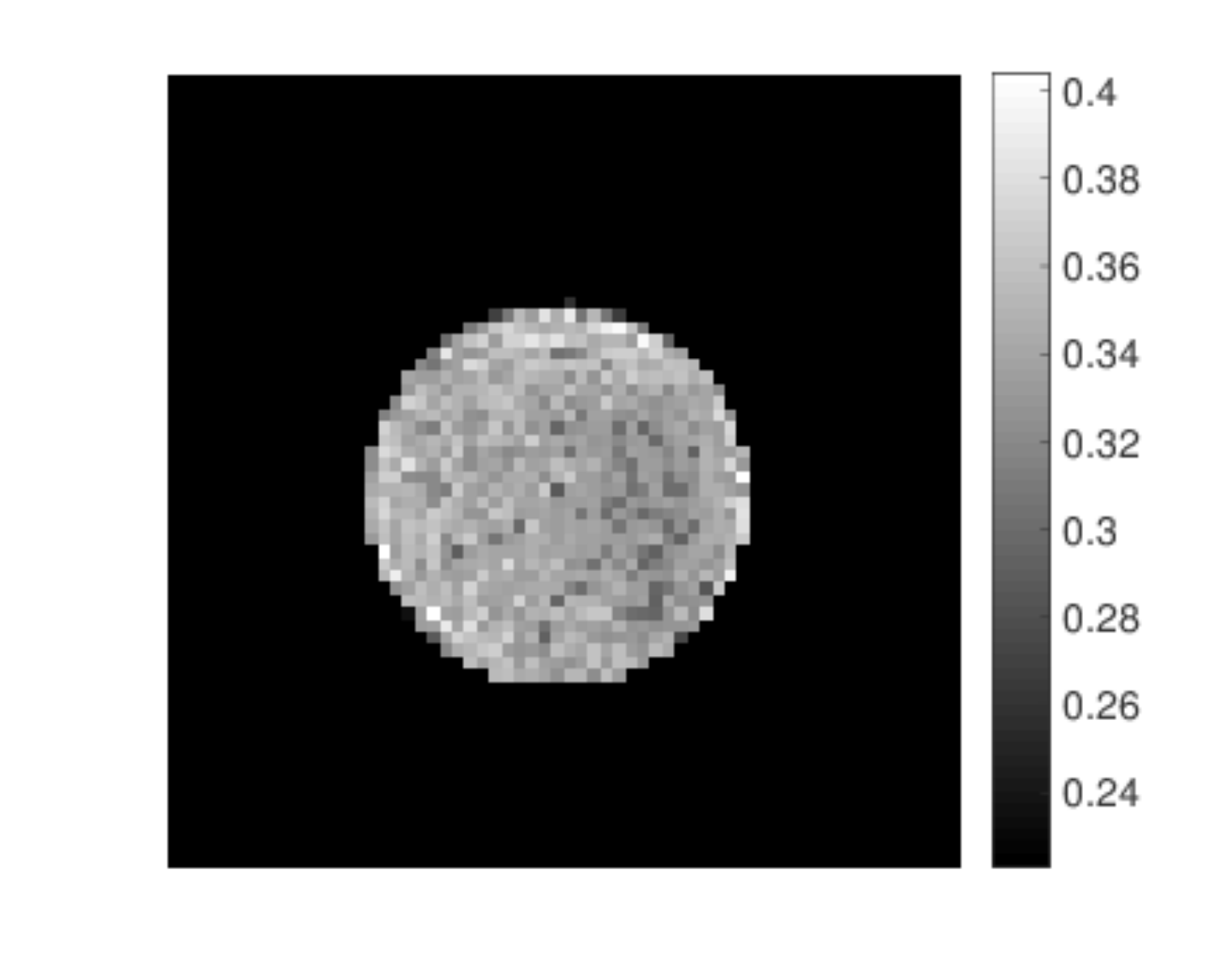}}}}

  \subfloat[$D^{(1)}$ Histogram]{\scalebox{0.4}{\adjustbox{trim={0.05\width} {0.0\height} {0.05\width} {0.05\height},clip}{%
          \includegraphics{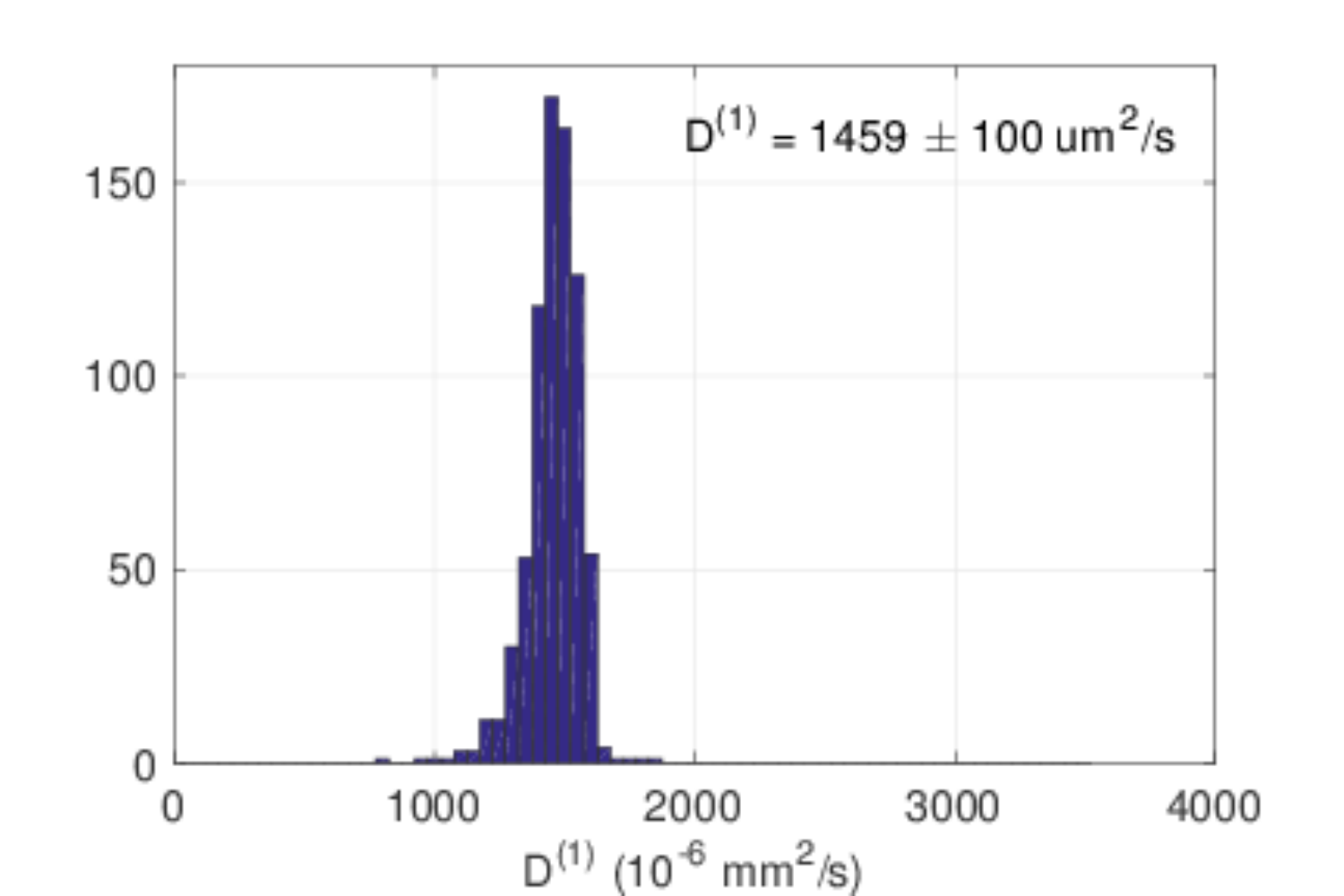}}}}
  \subfloat[$D^{(2)}$ Histogram]{\scalebox{0.4}{\adjustbox{trim={0.05\width} {0.0\height} {0.05\width} {0.05\height},clip}{%
          \includegraphics{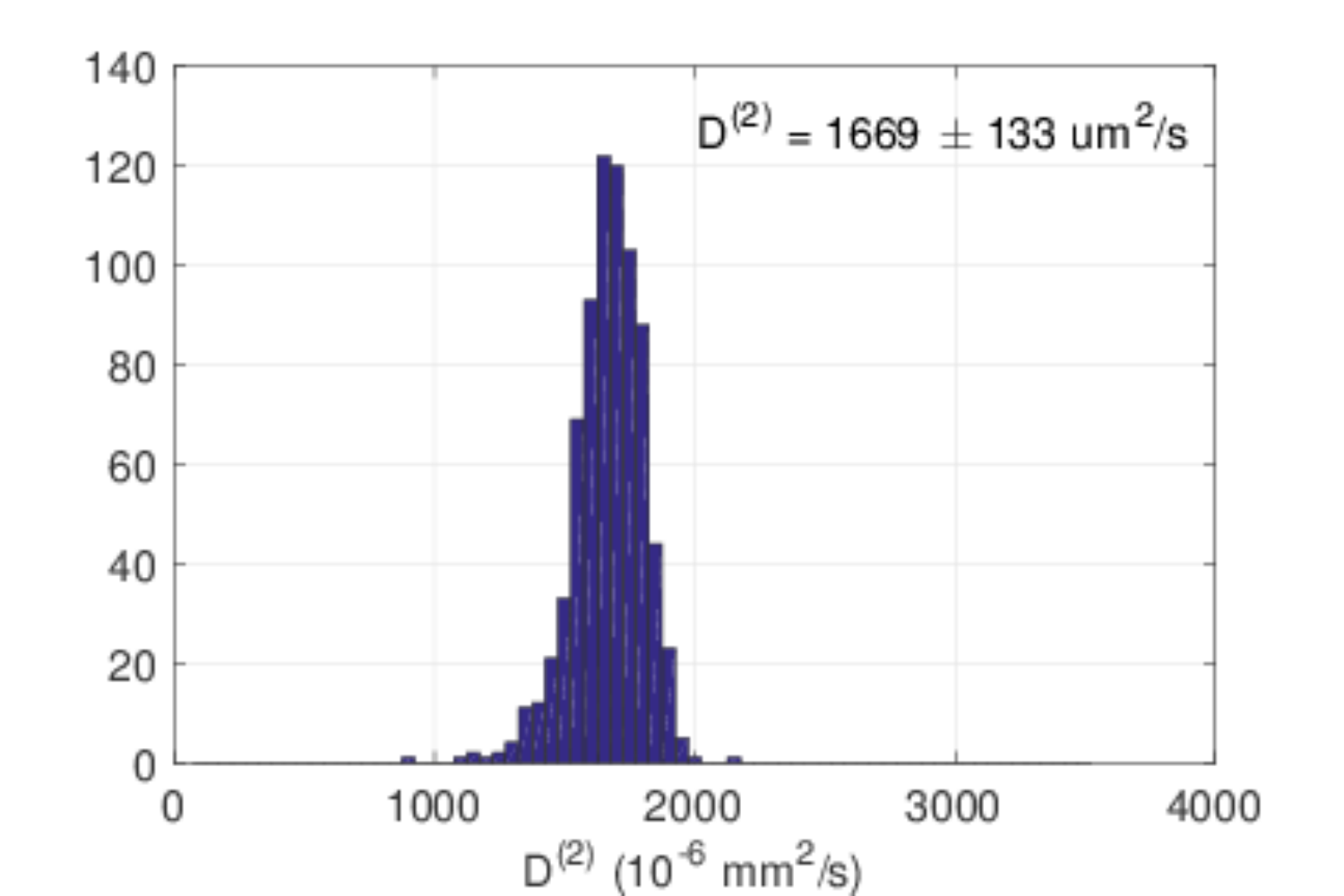}}}}
  \subfloat[Kurtosis Histogram]{\scalebox{0.4}{\adjustbox{trim={0.05\width} {0.0\height} {0.05\width} {0.05\height},clip}{%
        \includegraphics{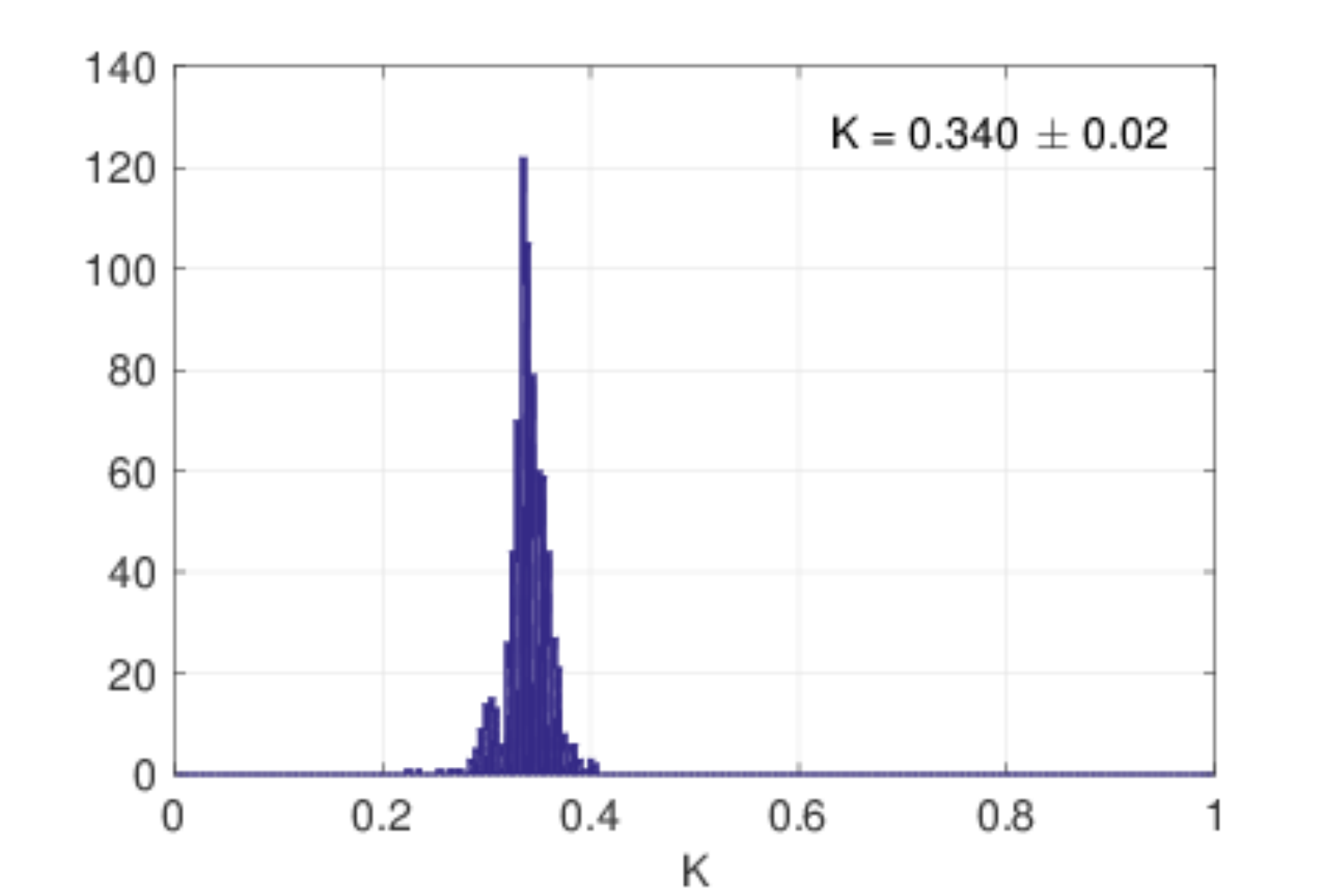}}}}
  \caption{Voxel-based analysis of diffusion data for a 2\% agar gel with \SI{2}{g} of microspheres. The top row shows the spatial distribution of the diffusion coefficients (a,b) and kurtosis (c).  The histograms (d,e,f)  show the corresponding statistical distribution of the values, indicating a spatial variation of the parameters as quantified by the standard deviation/mean of the distribution of 6-8\%.}
  \label{fig:maps+hist}

  \subfloat[$R_1$ Map ($1/s$)]{\!\!\!\!\!\!\!\scalebox{0.45}{\adjustbox{trim={0.07\width} {0.1\height} {0.03\width} {0.0\height},clip}{%
        \includegraphics{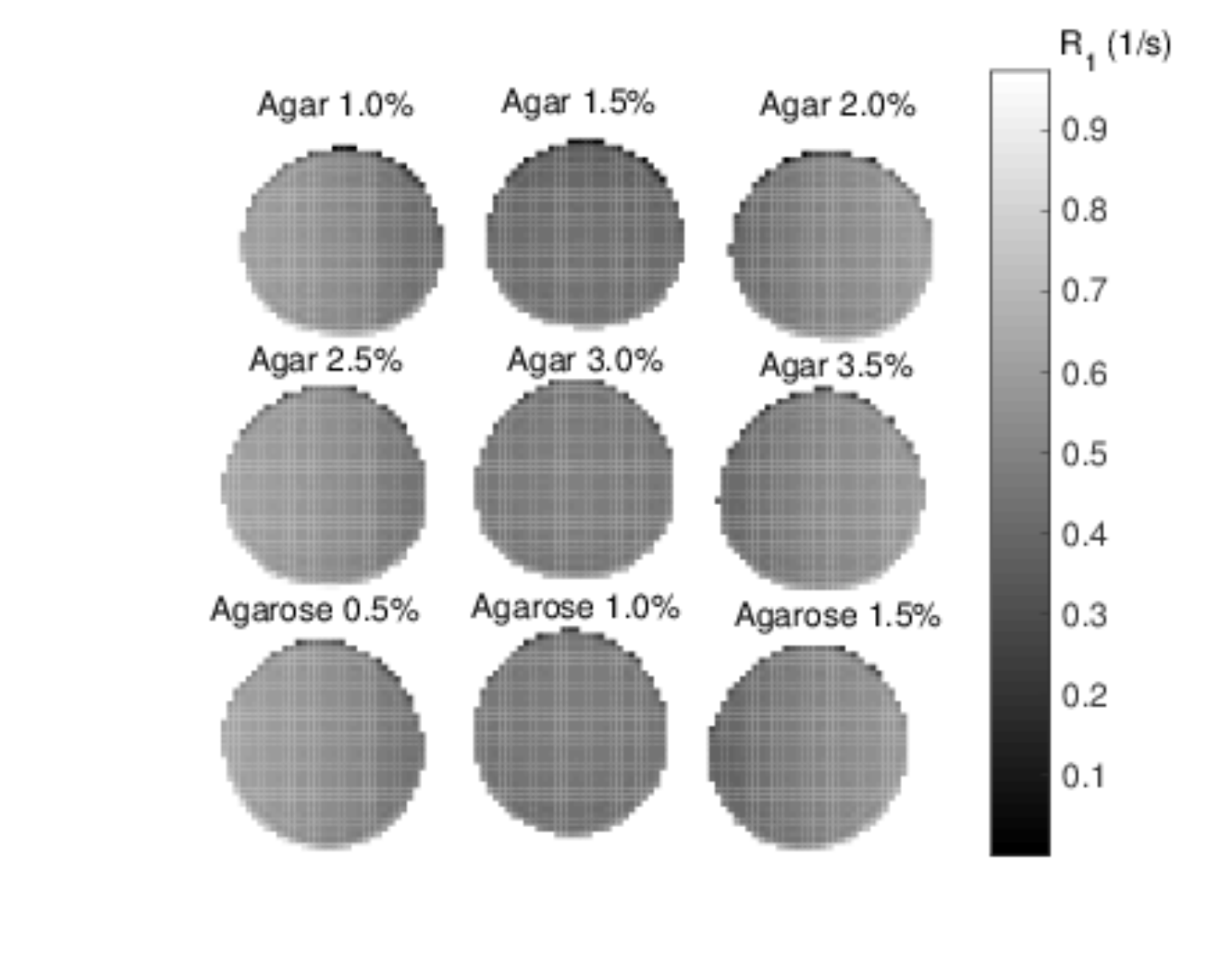}}}} 
  \subfloat[$R_1$ Map ($1/s$)]{\scalebox{0.45}{\adjustbox{trim={0.07\width} {0.1\height} {0.03\width} {0.0\height},clip}{%
        \includegraphics{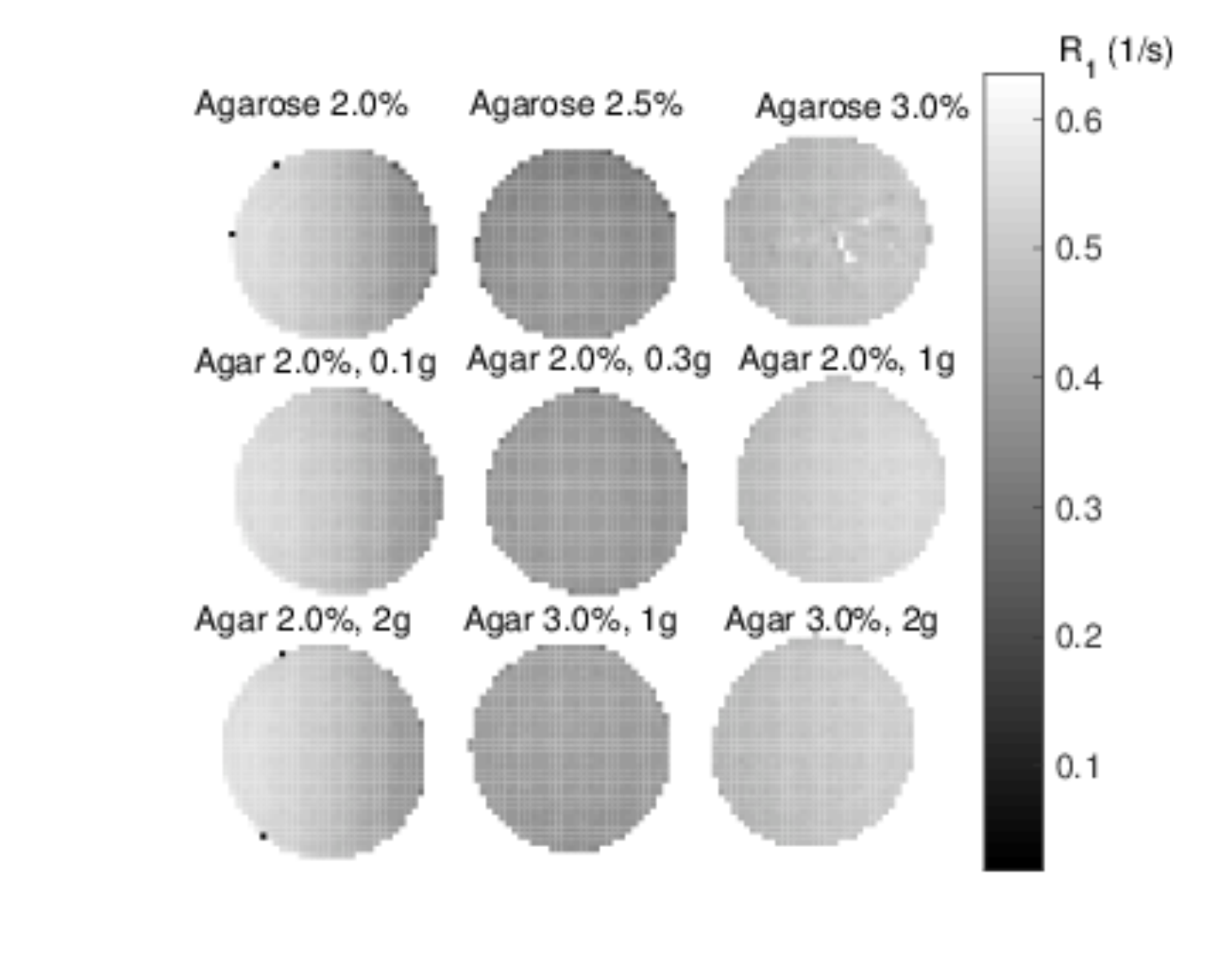}}}}
  \subfloat[$R_1$ Map ($1/s$)]{\scalebox{0.45}{\adjustbox{trim={0.07\width} {0.1\height} {0.03\width} {0.0\height},clip}{%
        \includegraphics{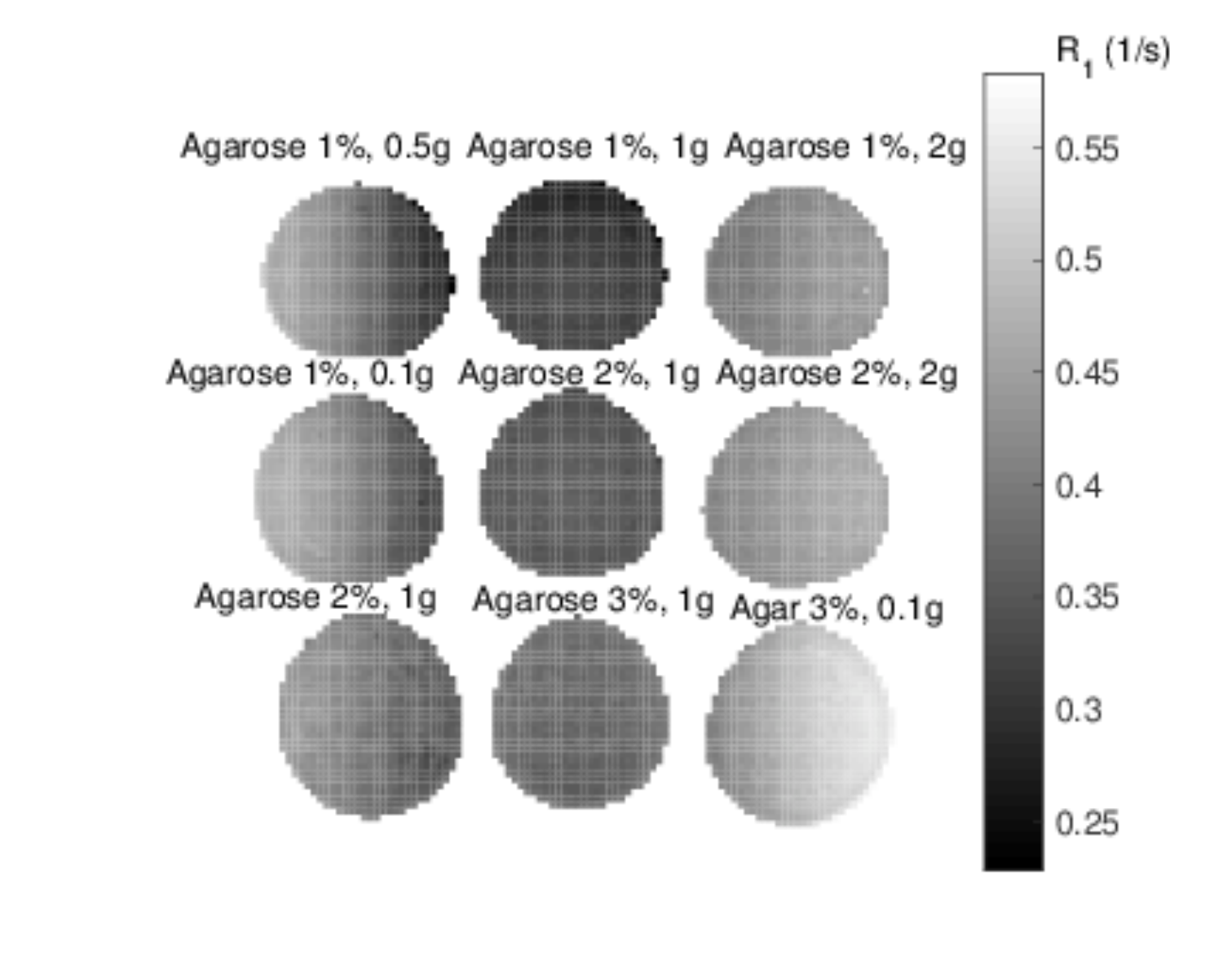}}}}\\ 
  \subfloat[$R_2$ Map ($1/s$)]{\!\!\!\!\!\!\scalebox{0.45}{\adjustbox{trim={0.07\width} {0.1\height} {0.03\width} {0.0\height},clip}{%
        \includegraphics{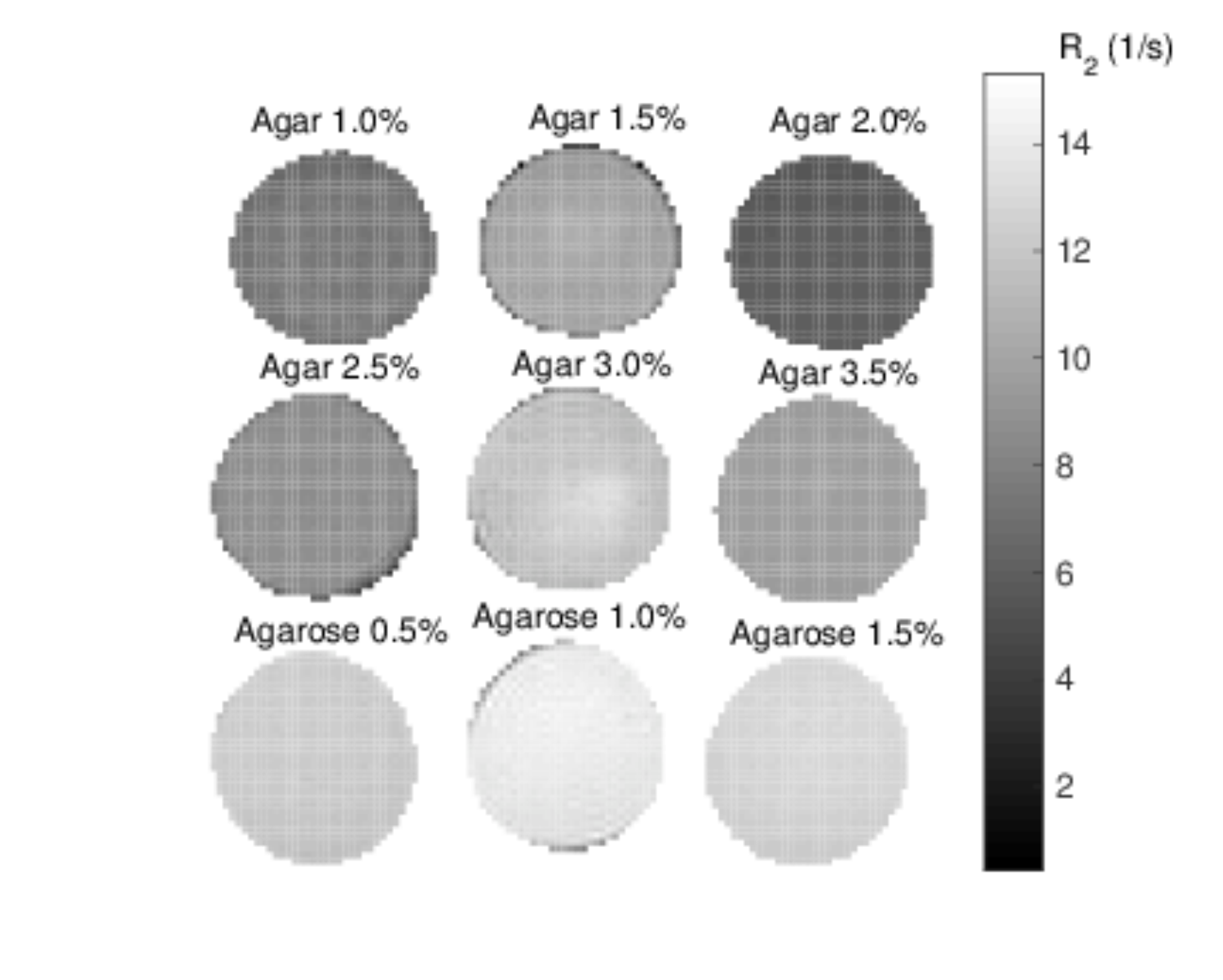}}}}
  \subfloat[$R_2$ Map ($1/s$)]{\scalebox{0.45}{\adjustbox{trim={0.07\width} {0.1\height} {0.03\width} {0.0\height},clip}{%
        \includegraphics{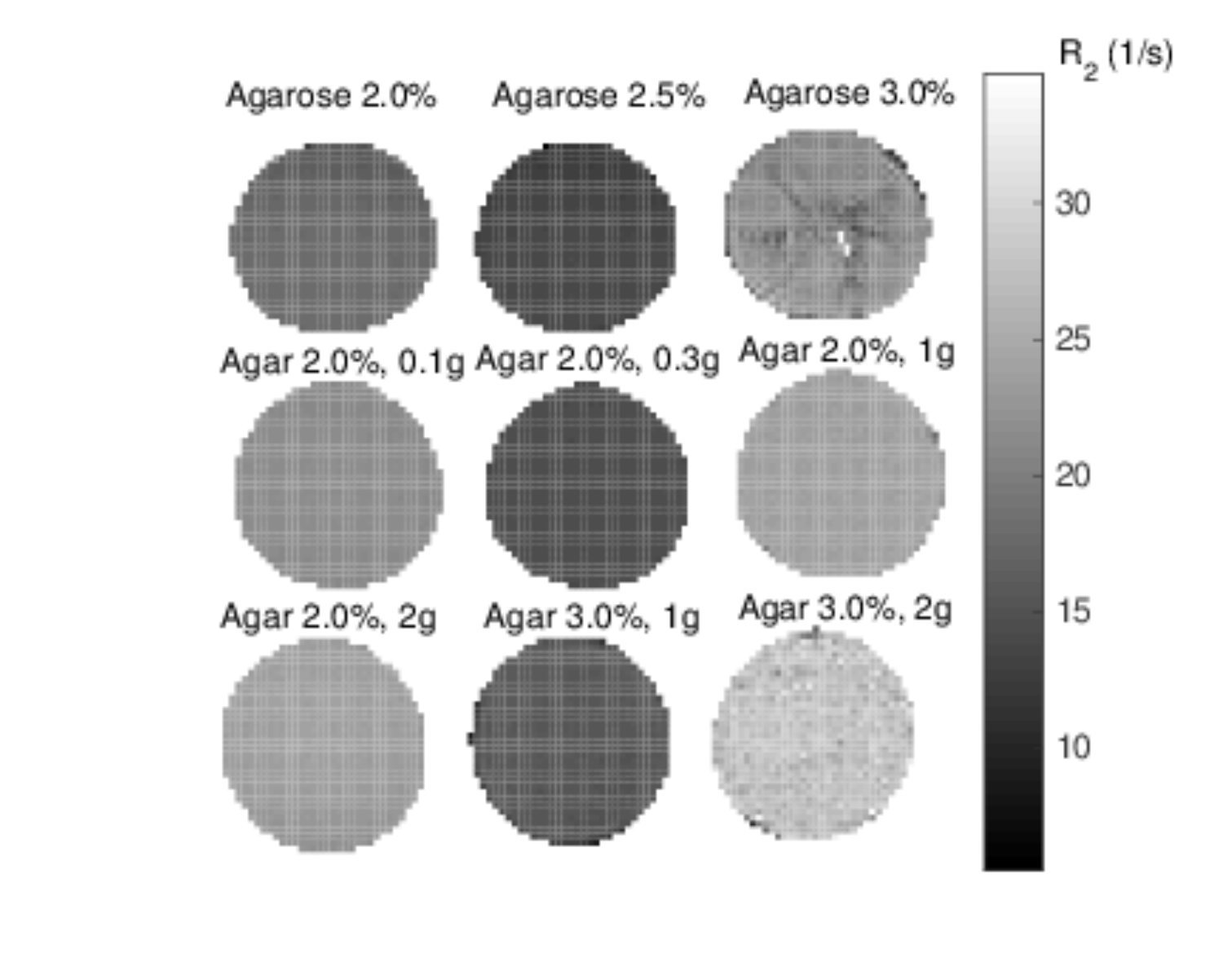}}}}
  \subfloat[$R_2$ Map ($1/s$)]{\scalebox{0.45}{\adjustbox{trim={0.07\width} {0.1\height} {0.03\width} {0.0\height},clip}{%
        \includegraphics{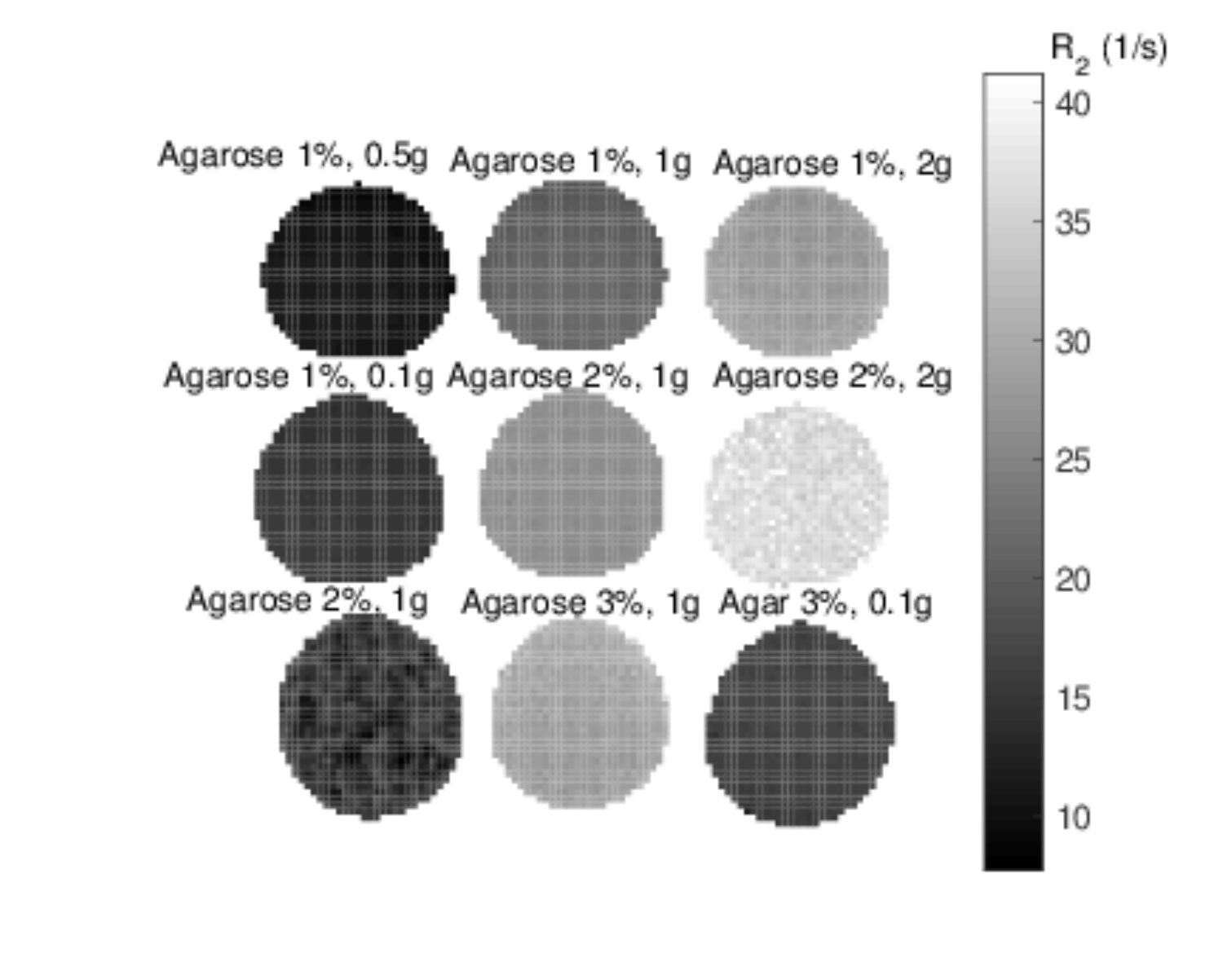}}}}\\
  \caption{$R_1$ (a)--(c) and $R_2$ (d)--(f) maps for 27 gels derived using voxel-based analysis.  Three series of scans were performed with 9 gels, arranged in a $3\times 3$ grid, scanned concurrently.   For all scans the same \SI{10}{mm} coronal slice through the middle of the phantoms was selected.}
  \label{fig:R-maps}
\end{figure*}

\enlargethispage{0.5in}
To assess the homogeneity of the phantoms additional multi-slice scans and single voxel analysis were performed (as detailed in the methods section).

Fig.~\ref{fig:R2s+Fieldmap}(a) shows the mean and variation of $R_2^*$ as a function of the slice index for series of \SI{2}{mm} coronal slices over a single phantom, for three different phantoms.  In some cases a substantial gradient in $R_2^*$ is observable, indicating possible concentration gradients (green curve) and for some gels an abrupt change in the $R_2^*$ indicates that the phantom has separated into layers (red curve).  These phantoms were excluded.  For a typical phantom (blue curve) $R_2^*$ changes only by a few percent, at least in the central part of the phantom.  For instance, in the difference between $R_2^*$ for slices 5 and 15 for the blue curve is about 8\%.  The larger variation of $R_2^*$ near the top and bottom of the phantom can likely be attributed to the greater $B_0$ inhomogeity in these regions.

To assess the $B_0$ homogeneity double-echo interference field maps were acquired for all phantoms.  Figure~\ref{fig:R2s+Fieldmap}(b) shows an example for a 2\% plain agar gel.  The distribution of the dark fringes indeed indicates greater inhomogeneity near the top and bottom of the phantom compared to the centre.  The same patterns were observed for both the pure gels and gels with microspheres.  To minimize the effects of field inhomogeneity all DKI and relaxation measurements were performed for a coronal slice through the centre, as detailed in the methods section.

To assess in plane homogeneity, a voxel-based analysis of the gels was performed.   Spatially resolved diffusion and kurtosis maps derived from voxel-based analysis of the data, shown for a 2\% agar gel with \SI{2}{g} of glass microspheres in Fig.~\ref{fig:maps+hist}, indicate some spatial variation but distributions for the diffusion and kurtosis parameters even for high concentrations of microspheres are still given by relatively narrow, approximately Gaussian distributions.

$R_1$ and $R_2$ maps for 27 gels were also acquired (see Fig.~\ref{fig:R-maps}).  Gels with high concentrations of microspheres generally show greater granularity but overall the $R_1$ and $R_2$ maps appear mostly uniform, as expected for homogeneous phantoms, with the variation, $\Delta R_1/\bar{R}_1$ and $\Delta R_2/\bar{R}_2$, observed for different phantoms ranging from 1.75\% to 16.5\%  and 0.72\% to 12.1\%, respectively.  Here $\bar{R}_n$ and $\Delta R_n$ denote the mean values and standard deviations of $R_n$ for $n=1,2$, observed for the distribution of values derived from the single voxel fits.  The fracture in the top right corner gel in Fig.~\ref{fig:R-maps}(b,e) is due to the phantom having been dropped accidentally.

To assess temporal stability of the gels, the $R_2$ scans were repeated after the gels had been stored at room temperature for over six months.  The resulting $R_2$ values can be found in Table~\ref{table:R}.

\begin{figure*}[!ht]
  \subfloat[Agarose 2\%, 2g micropheres]{\includegraphics[width=\columnwidth]{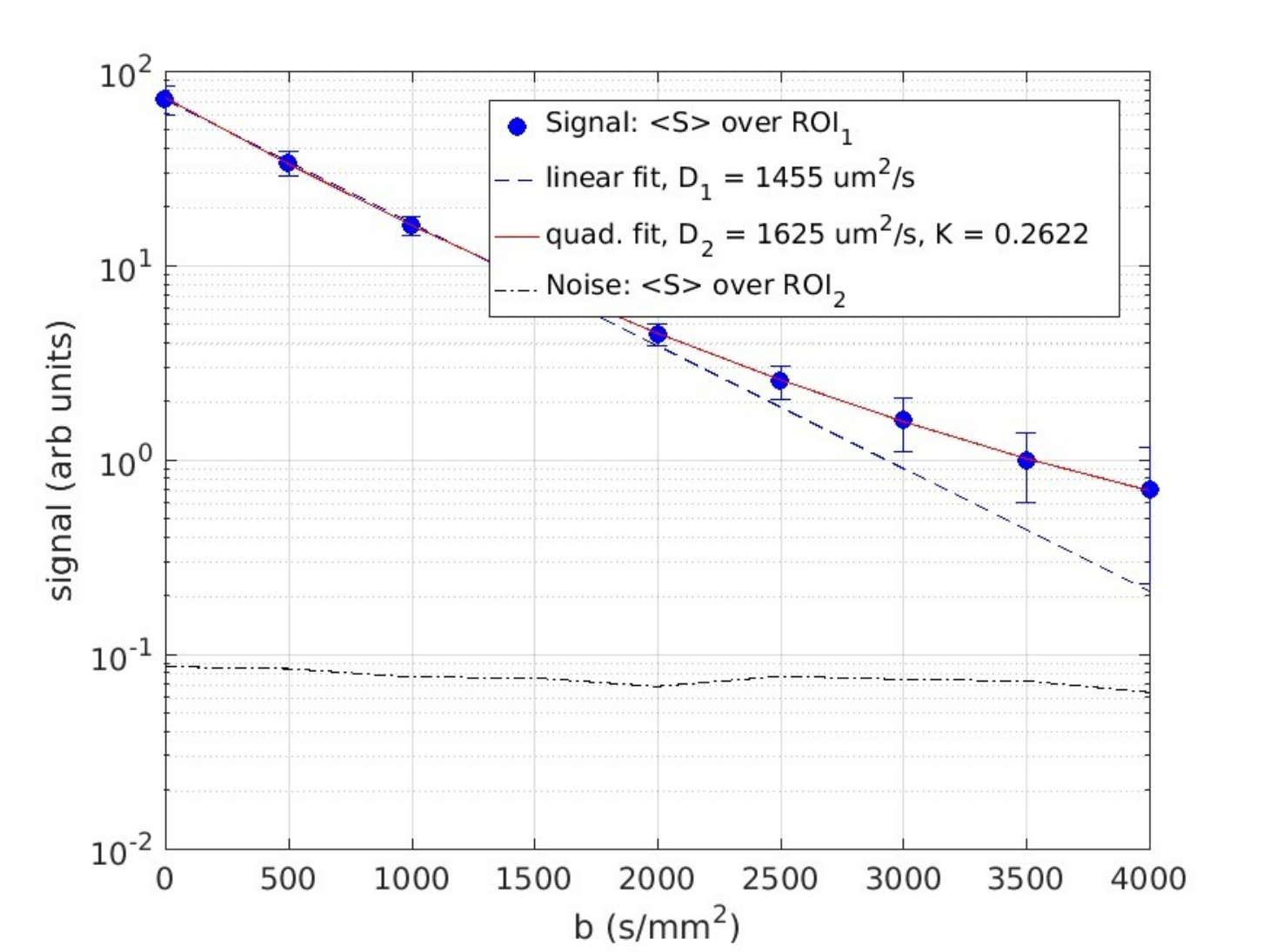}}
  \subfloat[Agarose 2\%, 2g micropheres]{\includegraphics[width=\columnwidth]{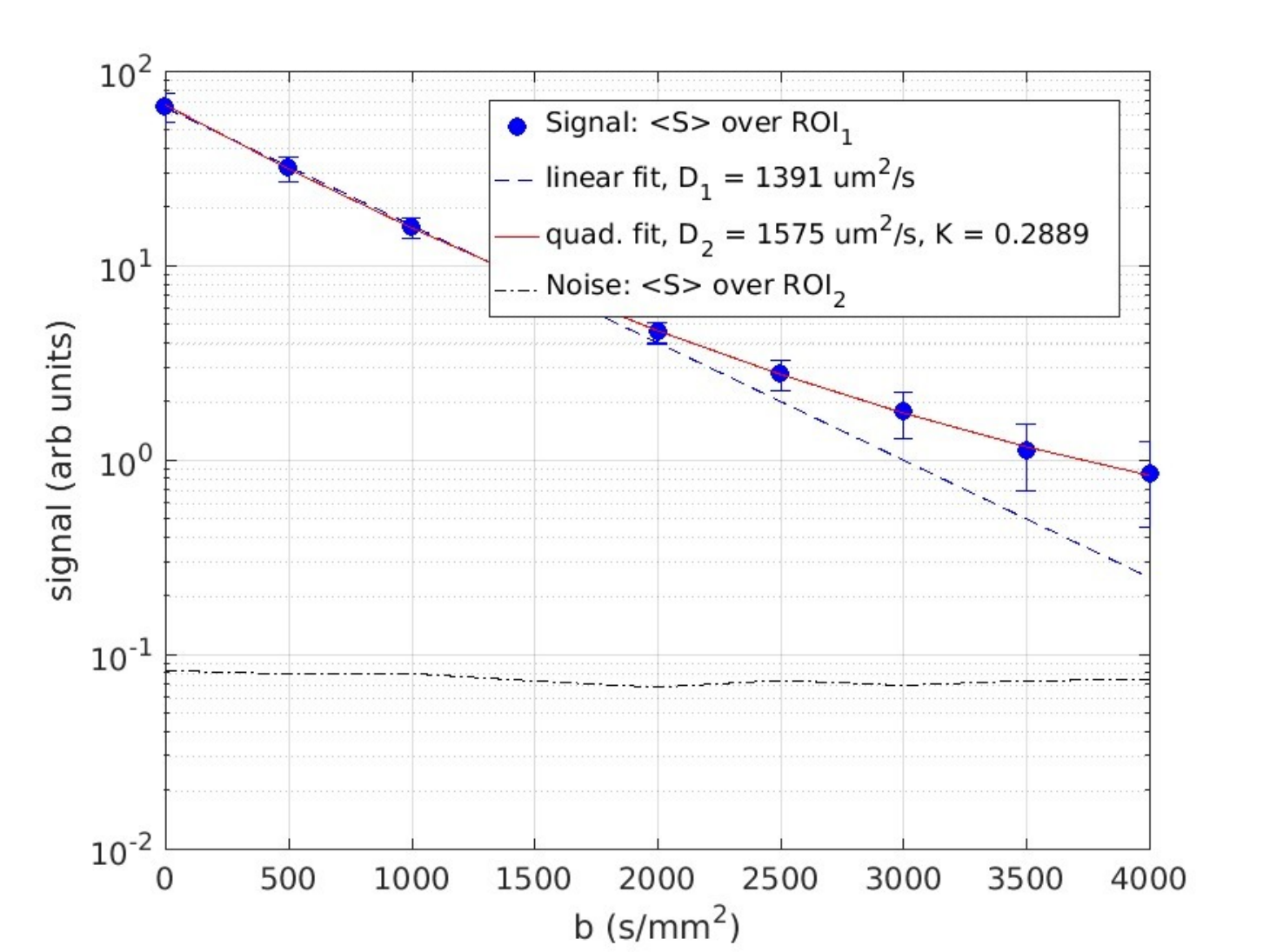}}
  \caption{The diffusion plots for two identically prepared agarose gels with the same concentration of the gelling agent and microspheres (acquired using identical scanning protocols).}
   \label{fig:Diffusion-reproducibility}
\end{figure*}

To assess the reproducibility of the results, multiple gels with the same composition were produced in a few cases.  Comparison of the results for two 2\% agarose gels with \SI{2}{g} microspheres in Fig.~\ref{fig:Diffusion-reproducibility} shows that the diffusion coefficients $D^{(1)}$ differ by 5\%, the diffusion coefficients $D^{(2)}$ by about 3\%, and the kurtosis values by about 10\%.  Thus, there is some variability but the results are reproducible to within a few percent, and the results could be further improved by enhanced process control of the preparation of the gels.

\subsection{Diffusion and Relaxometry Tables and Statistics}

\begin{table*}[!hbt]
 \scalebox{0.9}{\begin{minipage}{1.1\textwidth}
\begin{tabular}{|l||c|c|c|c|c|}
\hline
& $D^{(1)} (\SI{1e-6}{mm^2/s})$ &  $D^{(2)} (\SI{1e-6}{mm^2/s})$ & $K$            & PAC $b=\SI{0}{s/mm^2}$ & PAC $b=\SI{4000}{s/mm^2}$ \\\hline\hline
1.0\% Agar                       & 2184 (2140,2227) & 2288 (2198,2377) & 0.050 (0.029,0.071) & 200.70 &   3.41\\\hline
1.5\% Agar                       & 2133 (2090,2177) & 2308 (2196,2421) & 0.102 (0.081,0.123) & 141.86 & 10.44\\\hline
2.0\% Agar                       & 2134 (2106,2162) & 2303 (2212,2394) & 0.094 (0.076,0.111) & 183.17 &   5.27\\\hline
2.5\% Agar                       & 2121 (2060,2183) & 2329 (2223,2434) & 0.124 (0.106,0.141) & 124.96 &   8.29\\\hline
3.0\% Agar                       & 2078 (2048,2108) & 2318 (2161,2474) & 0.126 (0.100,0.152) & 121.12 &   4.67\\\hline
3.5\% Agar                       & 2055 (2016,2095) & 2694 (2210,3177) & 0.216 (0.185,0.246) &   56.20 &   1.12\\\hline
0.5\% Agarose                  & 2308 (2191,2425) & 2424 (2355,2513) & 0.060 (0.043,0.078) & 191.57 & 5.79\\\hline
1.0\% Agarose                  & 2259 (2168,2350) & 2344 (2283,2404) & 0.059 (0.046,0.072) & 136.97 & 4.32\\\hline
1.5\% Agarose                  & 2222 (2102,2341) & 2362 (2315,2409) & 0.089 (0.080,0.098) & 117.91 & 3.74\\\hline
2.0\% Agarose                  & 2186 (2007,2364) & 2355 (2261,2448) & 0.106 (0.089,0.122) & 105.79 & 2.63\\\hline
2.5\% Agarose                  & 2198 (1944,2451) & 2458 (2211,2706) & 0.137 (0.103,0.170) &   61.90 & 2.72\\\hline
3.0\% Agarose                  & 2028 (1956,2100) & 2267 (2088,2447) & 0.133 (0.103,0.164) &   61.28 & 3.47\\\hline
10\% PVA                          & 1855 (1827,1883) & 2210 (1995,2426) & 0.200 (0.172,0.228) & 193.28 & 35.35\\\hline
15\% PVA                          & 1631 (1606,1656) & 1890 (1740,2039) & 0.207 (0.177,0.237) & 173.51 & 34.53\\\hline
20\% PVA                          & 1421 (1400,1443) & 1586 (1484,1689) & 0.183 (0.146,0.219) & 153.06 & 38.20\\\hline
1.0\% Agar, 2.0g spheres    & 1372 (1023,1720) & 1573 (1443,1703) & 0.435 (0.415,0.456)& 75.72 & 24.70\\\hline
1.5\% Agar, 2.0g spheres    &    923 ( 782,1081) & 1030  ( 983,1077) & 0.449 (0.422,0.476)& 37.09 & 20.52\\\hline
2.0\% Agar, 0.1g spheres    & 2066 (2015,2117) & 2421 (2217,2625) & 0.177 (0.154,0.200)& 245.92& 6.96\\\hline
2.0\% Agar, 0.3g spheres    & 1945 (1808,2081) & 2215 (2142,2289) & 0.189 (0.179,0.199)& 138.68& 20.01\\\hline
2.0\% Agar, 1.0g spheres    & 1498 (1260,1737) & 1723 (1622,1823) & 0.330 (0.324,0.354)& 50.47  & 50.43\\\hline
2.0\% Agar, 2.0g spheres    &   986 ( 754,1218) & 1115 (1026,1205)  & 0.466 (0.431,0.501)& 16.56  & 16.32\\\hline
2.0\% Agar, 3.0g spheres    &   818 ( 585,1053) &   898 ( 819, 976)    & 0.523 (0.465,0.581)& 11.92  & 9.66\\\hline
2.5\% Agar, 2.0g spheres    & 1097 ( 621,1574) & 1202 (1023,1381)  & 0.518 (0.466,0.570)&  8.28   & 9.31\\\hline
3.0\% Agar, 0.05g spheres  & 2012 (1941,2082) & 2326 (2203,2449) & 0.190 (0.175,0.204)& 140.75& 8.43\\\hline
3.0\% Agar, 0.1g spheres    & 1913 (1812,2014) & 2200 (2123,2278) & 0.212 (0.203,0.222)& 239.82& 10.76\\\hline
3.0\% Agar, 1.0g spheres    & 1227 (1011,1443) & 1366 (1300,1431) & 0.351 (0.331,0.371)& 15.04  & 16.83\\\hline
3.0\% Agar, 2.0g spheres    &   979 ( 676,1283) & 1102 (1014,1190) & 0.523 (0.490,0.556) & 16.01  & 6.27\\\hline
3.0\% Agar, 3.0g spheres    & 1124 ( 920,1328) & 1231 (1069,1392) & 0.449 (0.398,0.501) & 43.41  & 2.28\\\hline
3.5\% Agar, 2.0g spheres    & 1082 ( 836,1328) & 1223 (1113,1332) & 0.476 (0.443,0.509) & 28.15  & 4.55\\\hline
1.0\% Agarose, 0.1g spheres & 2061 (1938,2184) & 2340 (2252,2429) & 0.168 (0.157,0.180) & 124.97 & 19.84\\\hline
1.0\% Agarose, 0.5g spheres & 2252 (2218,2286) & 2947 (2355,2739) & 0.126 (0.101,0.151) & 200.71 & 3.00\\\hline
1.0\% Agarose, 1.0g spheres & 1962 (1865,2060) & 2227 (2125,2328) & 0.196 (0.183,0.209) & 144.13 & 2.46\\\hline
1.0\% Agarose, 2.0g spheres & 2257 (2118,2296) & 2562 (2403,2721) & 0.127 (0.118,0.156) & 169.37 & 7.44\\\hline
2.0\% Agarose, 0.1g spheres & 2158 (2112,2205) & 2542 (2365,2719) & 0.182 (0.165,0.198) & 199.45 & 2.62\\\hline
2.0\% Agarose, 1.0g spheres & 1890 (1811,1969) & 2181 (2083,2278) & 0.203 (0.190,0.216) & 213.99 & 3.42\\\hline
2.0\% Agarose, 2.0g spheres & 1617 (1448,1785) & 1824 (1780,1868) & 0.262 (0.254,0.269) & 121.21 & 3.43\\\hline
3.0\% Agarose, 1.0g spheres & 1355 (1196,1513) & 1499 (1358,1640) & 0.289 (0.248,0.330) &   52.71 & 1.96\\\hline
\end{tabular}
\end{minipage}}
\caption{Diffusion and Kurtosis parameters and phantom-air contrast (PAC) at $b=0$ and $b=\SI{4000}{s/mm^2}$.
Numbers in parentheses indicate 95\% confidence interval for the parameter.}
\label{table:DWI}
\end{table*}

\begin{table*}[h!]
\scalebox{0.9}{\begin{minipage}{1.1\textwidth}
\begin{tabular}{|l||c|c|c|c|c|} \hline
Gel Phantom Material & Density (\si{kg/m^3}) & $R_1$ ($1/s$) & $R_2$  ($1/s$) 2016 & $R_2$ ($1/s$) 2017 & $R_2^*$ ($1/s$) \\\hline
1.0\% Agar        & 1016 & 0.36  (0.34, 0.38) &   6.34  ( 6.25,   6.43)  &  6.25  (  6.15,  6.35) &  21.89  ( 20.69,  23.10)\\\hline
1.5\% Agar        & 1021 & 0.39  (0.37, 0.41) &   6.26  ( 5.89,   6.64)  &  7.46  (  7.35,  7.58) &  25.12  ( 23.15,  27.10)\\\hline
2.0\% Agar        & 1026 & 0.40  (0.38, 0.42) & 11.67  (11.39, 11.95) & 11.54 (11.24, 11.85) &  29.93  ( 26.85,  33.00)\\\hline
2.5\% Agar        & 1031 & 0.41  (0.39, 0.43) &   8.37  (  8.17,   8.56) &  8.34  (  8.15,   8.53) &  29.94  ( 28.56,  31.33)\\\hline
3.0\% Agar        & 1036 & 0.42  (0.40, 0.43) & 10.68  (10.14, 11.21) & 10.36 (10.09, 10.64) &  34.27  ( 31.65,  36.89)\\\hline
3.5\% Agar        & 1041 & 0.44  (0.42, 0.47) & 13.78  (13.08, 14.48) & 12.89 (12.50, 13.29) &  38.38  ( 36.07,  40.69)\\\hline
0.5\% Agarose  & 1011 & 0.35  (0.33, 0.37) &   4.90  (  4.63,   5.16) &   4.98  (  4.89,  5.07)  &  24.68  ( 23.47,  25.88)\\\hline
1.0\% Agarose  & 1016 & 0.35  (0.33, 0.37) &   8.59  (  8.51,   8.67) &   8.76  (  8.61,  8.92)  &  27.27  ( 25.83,  28.71)\\\hline
1.5\% Agarose  & 1021 & 0.37  (0.35, 0.39) & 12.02  (11.52, 12.53) & 12.33  (11.97, 12.68) &  29.31  ( 27.22,  31.39)\\\hline
2.0\% Agarose  & 1026 & 0.39  (0.36, 0.42) & 15.69  (14.82, 16.55) & 15.82  (15.01, 16.62) &  35.11  ( 32.65,  37.57)\\\hline
2.5\% Agarose  & 1031 & 0.39  (0.37, 0.42) & 18.84  (17.38, 20.29) & 19.81  (18.14, 21.48) &  39.29  ( 36.63,  41.95)\\\hline
3.0\% Agarose  & 1036 & 0.39  (0.36, 0.42) & 21.36  (19.18, 23.54) & 22.14  (19.90, 24.39) &  40.58  ( 38.29,  42.88)\\\hline
10.0\% PVA 4ft  & 1056 & 0.69  (0.68, 0.69) &   6.18  (  6.11,   6.25) & ---                               & 23.64  ( 20.63,  26.64)\\\hline
15.0\% PVA 4ft  & 1066 & 0.89  (0.88, 0.90) & 11.04  (10.86, 11.22) & ---                               & 28.20  ( 25.51,  30.90)\\\hline
20.0\% PVA 4ft  & 1106 & 1.07  (1.05, 1.08) & 13.68  (13.20, 14.15) & ---                               & 32.63  ( 30.97,  34.30)\\\hline
1.0\% Agar, 2.0g spheres & 1036 & 0.45  (0.44, 0.46) & 13.31  (12.70, 13.92) & 10.41 ( 9.93,10.88)  &   21.89  ( 20.69,     23.10)\\\hline 
1.5\% Agar, 2.0g spheres & 1041 & 0.43  (0.41, 0.44)  & 16.16  (15.60, 16.72) & 17.34 (16.79, 17.90) & 233.80  (229.90, 237.70)\\\hline 
2.0\% Agar, 0.1g spheres & 1027 & 0.41  (0.40, 0.43)  & 12.36  (11.90, 12.83) & 11.74 (11.51, 11.97) &   55.01  ( 52.45,    57.57)\\\hline 
2.0\% Agar, 0.3g spheres & 1029 & 0.38  (0.36, 0.40)  & 12.17  (11.70, 12.64) & 12.66 (12.35, 12.96) &   37.60  ( 35.20,    40.00)\\\hline 
2.0\% Agar, 1.0g spheres & 1036 & 0.43  (0.42, 0.44) & 11.57  (11.15, 11.98) & 11.97 (11.65, 12.29) & 128.60  (126.10, 131.00)\\\hline 
2.0\% Agar, 2.0g spheres & 1046 & 0.46  (0.45, 0.47)  & 16.83  (16.01, 17.64) & 17.38 (16.03, 18.73) & 232.80  (222.10, 243.40)\\\hline 
2.0\% Agar, 3.0g spheres & 1056 & 0.47  (0.46, 0.48) & 17.30  (16.04, 18.55) & 18.60 (17.38, 19.81) & 292.50  (288.40, 296.50)\\\hline 
2.5\% Agar, 2.0g spheres & 1051 & 0.48  (0.47, 0.49) & 22.16  (20.33, 23.98) & 21.75 (20.12, 23.37) & 246.50  (232.50, 260.40)\\\hline 
3.0\% Agar, 0.05g spheres& 1036 & 0.41  (0.37, 0.43)  & 13.17 (12.54, 13.79) & 14.42 (13.68, 15.16) &   32.29  (  29.93,   34.65)\\\hline 
3.0\% Agar, 0.1g spheres  & 1037 & 0.41  (0.39, 0.43) & 13.22  (12.59, 13.85  & 14.05 (13.38, 14.72) &   39.67  (  37.42,   41.92)\\\hline 
3.0\% Agar, 1.0g spheres  & 1046 & 0.46  (0.44, 0.48)  & 20.66  (19.72, 21.59) & 21.93 (20.76, 23.11) & 129.50  (125.80, 133.10)\\\hline 
3.0\% Agar, 2.0g spheres  & 1056 & 0.51  (0.50, 0.52)  & 26.44  (23.34, 29.54) & 24.52 (22.57, 26.47) & 241.20  (236.50, 246.00)\\\hline 

1.0\% Agarose, 0.1g spheres & 1017 & 0.36  (0.35, 0.38) &   9.69  (  9.34, 10.04) & 10.12 (  9.63,10.6)   &  39.27 ( 33.51,  45.03)\\\hline 
1.0\% Agarose, 0.5g spheres & 1021 & 0.34  (0.32, 0.35) & 10.72  (10.32, 11.13) & 11.18 (10.65, 11.72) &  22.29 ( 17.71, 26.87)\\\hline 
1.0\% Agarose, 1.0g spheres & 1026 & 0.34  (0.35, 0.35) & 12.23  (11.76, 12.71) & 13.22 (12.64, 13.80) &  28.81 ( 24.89, 32.72)\\\hline 
1.0\% Agarose, 2.0g spheres & 1036 & 0.35  (0.34, 0.36) & 14.36  (14.10, 14.62) & 14.47 (13.74, 15.20) &  97.72 ( 89.04, 106.40)\\\hline 
2.0\% Agarose, 1.0g spheres & 1036 & 0.36  (0.34, 0.37) & 21.94  (19.90, 23.98) & 24.14 (21.13, 27.15) &  49.31 ( 45.09,  53.53)\\\hline 
2.0\% Agarose, 2.0g spheres & 1046 & 0.37  (0.36, 0.38) & 23.05  (21.05, 25.05) & 24.56 (22.79, 26.33) & 103.80  (102.90, 104.70)\\\hline 
3.0\% Agarose, 1.0g spheres & 1046 & 0.40  (0.39, 0.41) & 33.07  (27.10, 39.04) & 33.47 (27.82, 39.13) & 125.50  (121.40, 129.50)\\\hline 
\end{tabular}
\end{minipage}}
\caption{Density, $R_1$ and $R_2$ and $R_2^*$.   Values in parentheses indicate 95\% confidence interval for the respective parameter based on the available data.}
\label{table:R}
\end{table*}


\end{document}